\documentclass[fleqn,usenatbib,useAMS]{mnras}

\usepackage{newtxtext,newtxmath}  
\usepackage{amsmath}  
\usepackage{color}
\usepackage{xcolor}
\usepackage[normalem]{ulem}
\usepackage{graphicx}
\usepackage{float}
\usepackage{stfloats}
\usepackage{placeins}
\usepackage{subcaption}
\usepackage{booktabs}
\usepackage{multirow}
\usepackage{colortbl}
\usepackage{multicol}
\usepackage{textcomp}
\usepackage{changepage}
\usepackage{siunitx}  
\usepackage{enumitem}
\usepackage{gensymb}  
\usepackage{upgreek}  
\usepackage{newunicodechar}
\newunicodechar{−}{\ensuremath{-}}

\DeclareRobustCommand{\VAN}[3]{#2}
\let\VANthebibliography\thebibliography
\def\thebibliography{\DeclareRobustCommand{\VAN}[3]{##3}\VANthebibliography}

\usepackage[T1]{fontenc}

\title[Scaling wide-field imaging with LOFAR]{Scalable and robust wide-field facet calibration with LOFAR's longest baselines}

\author[J.M.G.H.J. de Jong et al.]{
J.M.G.H.J. de Jong$^{1}$,
L. Veefkind$^{2}$,
R.J. van Weeren$^{1}$,
J.B.R. Oonk$^{1,2}$,
R.J. Schlimbach$^{2}$,
D.N.G. Kampert$^{2}$,
\newauthor
M. van der Wild$^{3}$,
L.K. Morabito$^{3,4}$,
F. Sweijen$^{3}$,
A.R. Offringa$^{5,6}$,
H.J.A. Röttgering$^{1}$
\\
$^{1}$Leiden Observatory, Leiden University, PO Box 9513, 2300 RA Leiden, The Netherlands\\
$^{2}$SURF, Science Park 140, 1098 XG Amsterdam, The Netherlands\\
$^{3}$Centre for Extragalactic Astronomy, Department of Physics, Durham University, Durham DH1 3LE, UK\\
$^{4}$Institute for Computational Cosmology, Department of Physics, Durham University, South Road, Durham DH1 3LE, UK\\
$^{5}$ASTRON, The Netherlands Institute for Radio Astronomy, Postbus 2, 7990 AA Dwingeloo, The Netherlands\\
$^{6}$Kapteyn Astronomical Institute, P.O. Box 800, 9700 AV Groningen, The Netherlands
}

\date{Accepted XXX. Received YYY}

\pubyear{\the\year{}}

\begin{document}

\label{firstpage}
\pagerange{\pageref{firstpage}--\pageref{lastpage}}
\maketitle

\begin{abstract} 
Recent work has successfully achieved sub-arcsecond wide-field imaging with high-band observations from the Low Frequency Array (LOFAR). However, the scalability of this work remains limited due to the need for manual intervention, poor calibration solutions for the Dutch LOFAR stations, and high computational costs.
We address these issues by: (1) improving automated self-calibration using a signal-to-noise metric and a neural network for image artefact detection; (2) implementing a refined calibration strategy for the Dutch LOFAR stations; and (3) cutting computational costs by optimising the data processing strategy.
We demonstrate the effectiveness of our automated processing strategy by reprocessing one previously reduced dataset and a new dataset from the ELAIS-N1 deep field, which features more severe ionospheric conditions. 
We find calibration artefacts across facet boundaries to be reduced with our improved automated calibration strategy and achieve a computational cost reduction of about a factor of 4 to 6 compared to previous work, where the exact factor depends on whether a single observation is processed or multiple observations of the same sky area are combined.
Further optimisation and improved handling of data with baseline-dependent averaging could reduce this in the near future by another factor of two, bringing the total cost for an 8-hour observation below 30,000 CPU core hours.
This work enables ultra-deep imaging at sensitivities on the order of a few $\upmu$Jy~beam$^{-1}$. Furthermore, it also lays the foundation for a fully automated survey pipeline for sub-arcsecond wide-field imaging of the northern sky with LOFAR.
\end{abstract}

\begin{keywords}
radio continuum: general --
surveys --
instrumentation: interferometers --
software: development
\end{keywords}

\section{Introduction}\label{sec:introduction}

The LOw Frequency ARray \citep[LOFAR;][]{haarlem2013} is one of the leading instruments to conduct deep high-angular resolution wide-field surveys at the lowest frequencies. LOFAR is sensitive to detecting low-frequency radio waves between 10-80~MHz with the low-band antennas (LBAs) and between 110-240~MHz with the high-band antennas (HBAs). LOFAR consists of stations distributed across the European continent, with its core in the Netherlands. When combined, LOFAR offer baselines of up to 2,000~km across Europe, enabling sub-arcsecond resolution wide-field imaging with field of views of about 2.5$\times$2.5~deg$^{2}$ with the HBAs \citep{sweijen2022}. These resolutions enable a broad range of scientific opportunities \citep[e.g.][]{morabito2025a}. Nonetheless, the scalability of calibrating and imaging observations with HBAs from all LOFAR stations for deep wide-field imaging is currently constrained by the high computational costs and challenges in achieving high-quality calibration solutions for both short and long baselines simultaneously \citep{dejong2024}.

Efforts to automatically calibrate and image LOFAR observations with only the Dutch LOFAR HBAs have led to the LOFAR Two-metre Sky Survey \citep[LoTSS;][]{shimwell2017, mechev2017, mechev2018, shimwell2022, williams2019} and the LoTSS-Deep Fields \citep{tasse2021, sabater2021, bondi2024, shimwell2025}.
The wide-field survey component provides images of the Northern sky at 144~MHz with 6\arcsec~resolution and median sensitivities of about \textasciitilde80~$\upmu$Jy~beam$^{-1}$. With 14 million sources detected across 89\% of the Northern sky Shimwell et al. (in prep.), this represents the largest radio survey conducted to date. The state-of-the-art for the Deep Fields is created with \textasciitilde500~hours of observations of the ELAIS-N1 deep field, reaching a central sensitivity of about 11~$\upmu$Jy~beam$^{-1}$ \citep{shimwell2025}. However, approximately 40\% of the noise in this image arises from confusion noise, highlighting the sensitivity limitations of imaging with only the Dutch LOFAR stations. Increasing the angular resolution of the observations, by incorporating all LOFAR stations across Europe (Dutch and international), is the most effective way to mitigate confusion noise. This demands the processing of larger data volumes, which are about 16~TB for an uncompressed 8-hour LOFAR observation, and therefore requires a data reduction pipeline capable of calibrating and imaging such large datasets efficiently.

Building on earlier successful work to process LOFAR data for sub-arcsecond resolution imaging \citep[e.g.][]{moldon2015, varenius2015, varenius2016, jackson2016, harris2019}, the next step to advance the development of a sub-arcsecond imaging strategy for LOFAR was introduced by \cite{morabito2022}, who focused on the direction-independent (DI) calibration of the full European array, followed by postage stamp imaging of targets. \cite{sweijen2022} extended this work to wide-field imaging with improved direction-dependent (DD) calibration, producing the first image that captured thousands of sources within a 2.5$\times$2.5~deg$^{2}$ field of view at a resolution of 0.3\arcsec. This was followed by \cite{dejong2024}, who imaged datasets from four observations of the same pointing of the ELAIS-N1 deep field, resulting in the detection of four times more sources within the same sky area compared to the imaging of only one observation by \cite{sweijen2022}. In addition to increasing depth, they introduced a way to perform automated DD calibrator selection, aiming to reduce the need for manual intervention. They also demonstrated that the current DD calibration strategy required additional attention since the calibration solutions, corresponding to the Dutch LOFAR stations and hence the shorter baselines, were of insufficient quality to enable high-resolution imaging. This issue was partly mitigated by introducing an additional ad hoc calibration step specifically for the Dutch LOFAR stations after completing all other calibration steps. While this solution reduced the problem, image artefacts across different facets of the images persist and therefore a refinement of the calibration strategy is required. Moreover, \cite{sweijen2022}, \cite{ye2024}, and \cite{dejong2024} demonstrated that the final subtraction and imaging steps consume the majority of data processing costs, accounting for approximately 80\% of the total expenses. Revising key aspects of the wide-field imaging strategy is therefore essential before scaling up to imaging for deeper or wider imaging.

In this work, we revisit and automate the DD calibration and develop strategies to reduce the required computational resources for wide-field sub-arcsecond imaging. We demonstrate the new calibration and imaging strategy on two LOFAR datasets from the ELAIS-N1 field, of which one has already been reduced by \cite{dejong2024} and one new dataset with more challenging ionospheric conditions.
With hundreds of hours of data available for this field recorded with Dutch and international LOFAR stations, we pave the way for ultra-deep imaging of a single pointing, reaching a RMS noise on the order of a few $\upmu$Jy~beam$^{-1}$ at 144~MHz.
Furthermore, our optimised DD calibration workflow can be directly integrated into a surveys pipeline for high-resolution imaging.

We start in Section \ref{sec:data} with a brief description of the test data that we used in this work.
In Section \ref{sec:automated}, we describe ways to further automate the self-calibration of LOFAR long-baseline data, setting the stage for the optimised calibration strategy presented in Section \ref{sec:ddcal}. In Section \ref{sec:datavolcomp}, we introduce methods for reducing data volumes. Section \ref{sec:results} presents our improved results, which are discussed along with future prospects in Section \ref{sec:discussion}. Finally, we end our work with conclusions in Section \ref{sec:conclusion}.

\section{Data}\label{sec:data}

For the purpose of this work, we have selected two datasets from the ELAIS-N1 field to serve as test cases for our upgraded data reduction strategy. ELAIS-N1 is a famous deep field, which has been explored across various wavelengths \citep[e.g.][]{manners2003, martin2005, lawrence2007, sirothia2009, mauduit2012, kondapally2021}. This extensive multi-wavelength coverage has established ELAIS-N1 as an important field for extragalactic studies. With the work done by \cite{ye2024} and \cite{dejong2024} to reduce LOFAR data for sub-arcsecond resolution wide-field imaging for this field, we have an advantage in experimenting with different calibrator sources and automated imaging settings to enhance already existing images. With over 500~hours of LOFAR observations available for ELAIS-N1 \citep{shimwell2025}, this field also serves as a strong candidate for ultra-deep imaging.

\begin{table*}
\caption{Metadata from the two ELAIS-N1 observations used in this work. Observations used calibrator 3C~295 as the primary calibrator. The number of stations and frequencies are recorded after flagging. The number of non-Dutch stations is indicated between brackets.}
\centering
\begin{tabular}{l l l l l l l l} \toprule
\textbf{Observ. ID} & \textbf{Project} & \textbf{Calibrator ID} & \textbf{Observ. date} & \textbf{Pointing centres} & \textbf{Stations (int)} & \textbf{Frequencies} \\ \midrule

\textbf{\texttt{L686962}} & \texttt{LT10\_012} & \texttt{\texttt{L686958}} & 26-11-2018 & 16:11:00, +54.57.00 & 49 (11) & 120-166~MHz \\

\textbf{\texttt{L833466}} & \texttt{LT14\_003} & \texttt{L833474} & 09-10-2021 & 16:11:00, +55.00.00 & 50 (13) & 118-166~MHz \\
   \bottomrule
\end{tabular}
\label{table:metadata}
\end{table*}

The datasets corresponding to the LOFAR observations are detailed in Table \ref{table:metadata}. The first dataset, \texttt{L686962}, is selected from project \texttt{LT10\_012}, representing an observation with a calm ionosphere and for use as a benchmark to validate the improved calibration and imaging strategies. We used the dataset already processed up to the in-field calibration. The second dataset, \texttt{L833466}, comes from project \texttt{LT14\_003} and features data recorded with a 2-sec integration time, compared to the default 1-sec, which leads to additional smearing at the edges \citep{dejong2024}. This observation was taken closer to solar maximum and shows stronger ionospheric effects compared to \texttt{L686962}. However, the initial calibration solutions from the LOFAR Initial Calibration (\texttt{LINC}) pipeline\footnote{\url{https://linc.readthedocs.io}} indicated that the dataset was still calibratable. This dataset serves to test the robustness of our new calibration strategy on challenging observations, while still remaining within the limits of what our methods can calibrate. By combining these two datasets, we also evaluate the effectiveness of joint-calibration across multiple observations (see Section \ref{sec:selfcal2}).

\section{Automated direction-dependent calibration}\label{sec:automated}

The ionosphere induces DD effects (DDEs) that lead to artefacts on scales from arcseconds to arcminutes \citep[e.g.][]{intema2009, smirnov2011}. To correct for this, we apply self-calibration on calibrators distributed across the field of view. These calibrators define the facets in a Voronoi tessellation, which is a division of the field into regions where each point is closest to a particular calibrator source, where the calibration solutions are assumed to be constant \citep{schwab1984, vanweeren2016}. We aim to achieve sub-arcsecond resolution images, which necessitates finding calibrators with high enough signal-to-noise ratios (S/N) at the longest baselines. This becomes more challenging at higher resolutions, as many sources with high S/N at lower resolutions become resolved out, leaving fewer sources with sufficient flux density at 0.3\arcsec~resolution.

In this section, we revisit the calibrator selection from \cite{dejong2024} and connect it to a metric for determining solution intervals, which improves automatic self-calibration. Additionally, we introduce a neural network to assess self-calibration convergence. This helps to refine the automation of calibration strategies by identifying the optimal number of self-calibration cycles and determining calibration parameters for each calibrator source without the need for human intervention. To enhance reproducibility, we have also added new functionalities to the \texttt{facetselfcal} calibration software package \citep{vanweeren2021}.\footnote{\url{https://github.com/rvweeren/lofar_facet_selfcal}}

\subsection{Calibrator selection}\label{sec:selection}

The first main step is to automatically identify high S/N calibrator sources across the field of view. \cite{dejong2024} implemented a source selection method that exploits the fact that circular polarisation is a rare phenomenon at low radio frequencies \cite[e.g.][]{callingham2023}. Therefore, the difference between right- and left-handed polarisation (RR-LL) should be minimal and the variance of the calibration corrections to correct for this effect should be small for high S/N calibrators.\footnote{This is a simplified assumption ignoring the effects of polarisation leakage variations and the minor impact of differential Faraday rotation for different directions.} To account for phase wrapping around $2\pi$, we use the circular standard deviation \citep[e.g.][]{mardia1972,fisher1993} on the calibration corrections for the phase difference between both polarisation directions. Since we are in particular interested in the S/N at the smallest angular scales, we only assess the circular standard deviation for the international stations, as these stations correspond to the longest baselines. In this work, we refer to this metric with the phasediff-score (denoted by $\hat{\sigma}_{c}$). In Appendix \ref{sec:phasediff_snr}, we provide statistical evidence that the phasediff-score is a valid proxy for S/N.

\begin{figure}
 \centering
\includegraphics[width=1\linewidth]{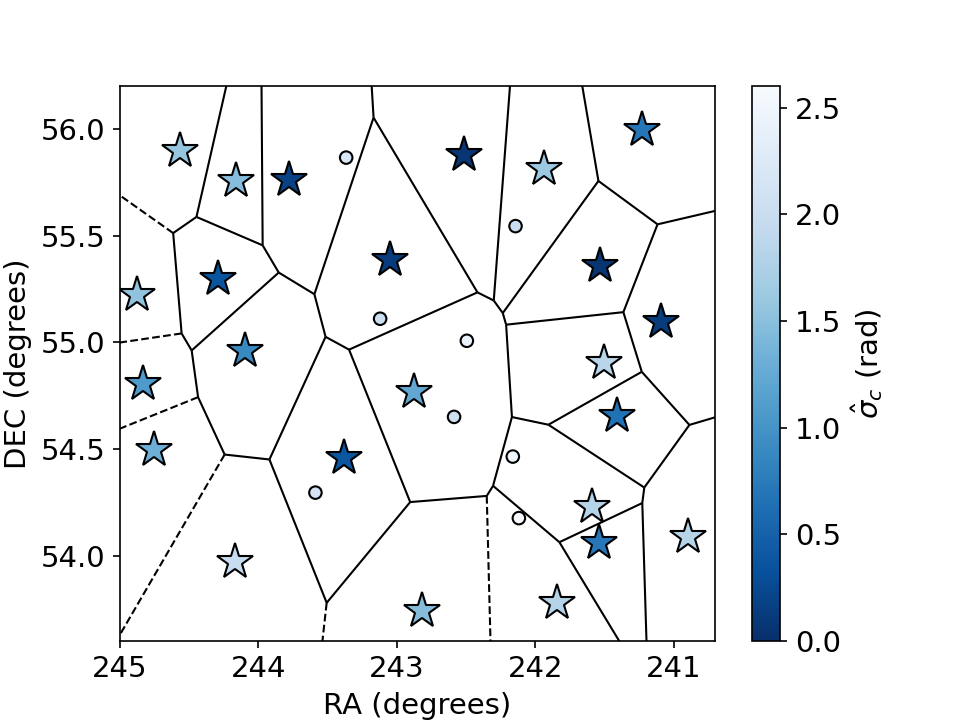}
  \caption{Main DD calibrator sources (stars) and weak facet calibrators (circles) with their respective phasediff-scores ($\sigma_{c}$) and corresponding facet boundaries in black.}
\label{fig:voronoi}
\end{figure}

To compare the phasediff-score of different sources, \cite{dejong2024} selected a fixed solution interval of $\hat{\delta}_{t}=10$~min for the calibration step. While they used only a phasediff-score threshold of $\hat{\sigma}_{c}<2.3$~rad, we opt to add a second group with lower S/N sources, which gives us the following two groups:
\begin{itemize}[leftmargin=*]
\item \textit{Main facet calibrators}: $\hat{\sigma}_{c}<2.3$~rad.
\item \textit{Weak facet calibrators}: $2.3\leq\hat{\sigma}_{c}<2.6$~rad.
\end{itemize}
The main facet calibrators define our facet layout (see Section \ref{sec:internationalcal} and Figure \ref{fig:voronoi}), while the second group consists of weaker `secondary' DD-calibrators, which can be used to refine DD calibration (see Section \ref{sec:selfcal2}). The option to incorporate weaker secondary sources to enhance image quality, by calibrating for more local DDEs after facet subtraction in sub-facets (see Section \ref{sec:subtract}), is particularly important for observations affected by a more turbulent ionosphere, where more calibrator sources are needed to obtain more local calibration corrections.

We also set in our source selection algorithm an additional minimal distance criterium between calibrator candidates to avoid having neighbouring facets with calibrator sources near the edge of the other facet, which may otherwise lead to DD artefacts crossing the facet boundary (this will be later demonstrated in an example in Section \ref{sec:facetbound}) or to having sources obtaining higher $\hat{\sigma}_{c}$ values due to picking up flux from bright neighbours. Since DDEs do not vary as much on small angular scales, we based our choice of a minimum distance value on the comparison between our two observations, where one has a wilder ionosphere than the other, and selected 0.15~deg as the minimum distance. If two or more sources are neighbouring within these distance thresholds, the source with the lowest phasediff-score is kept and the other one(s) is removed from the selection. Note that this value may still be adjusted as we continue processing datasets from different sky regions and under varying ionospheric conditions.

When combining both observations to compute the phasediff-scores, we find an average phasediff-score of 1.28~rad with a standard deviation of 0.72~rad. This combined selection is done by determining the phasediff-score on the calibration solutions from both observations. Since one of the two observations exhibits a more disturbed ionosphere than the other (see Section \ref{sec:data}), we also analysed the phasediff-score distributions separately for each observation. For the observation with a calmer ionosphere (\texttt{L686962}), the average phasediff-score for the main facet calibrators is 1.26~rad, with a standard deviation of 0.70~rad. The observation with a more turbulent ionosphere (\texttt{L833466}) shows a slightly higher average phasediff-score of 1.30~rad and a standard deviation of 0.75~rad. The small difference between the phasediff-scores of both observations aligns with the rationale presented in \cite{dejong2024}, where the metric based on the RR-LL polarisation difference was designed to cancel out the effects from ionospheric phase delays, thereby reducing sensitivity to ionospheric variability. We identify two calibrator sources with phasediff-scores slightly below 2.3~rad for observation \texttt{L686962} but above 2.3~rad for observation \texttt{L833466}. Both are located near facet boundaries, where smearing is more pronounced and the S/N is correspondingly lower. This may further increase the phasediff-scores for \texttt{L833466}, given that this dataset was recorded with double the time resolution of \texttt{L686962}.

For the calibrator selection for the ELAIS-N1 datasets used in this work, the algorithm removed two sources from the main facet calibrators and seven from the weak calibrators because they were located within 0.15~deg of another calibrator. The final selection yields 23 strong calibrators, along with the already calibrated in-field calibrator and 8 weak facet calibrators. The facet layout corresponding to the main DD-calibrators, along with the positions of the main and weak facet calibrators, is shown in Figure \ref{fig:voronoi}.

\subsection{Solution interval metric}\label{sec:solints}

Since the phasediff-scores are linked to the S/N of the longest baselines (see Appendix \ref{sec:phasediff_snr}), they also relate to the optimal width of the solution interval for self-calibration. This enables us to derive a metric for determining solution intervals for each calibrator source.

\begin{figure}
 \centering
\includegraphics[width=1\linewidth]{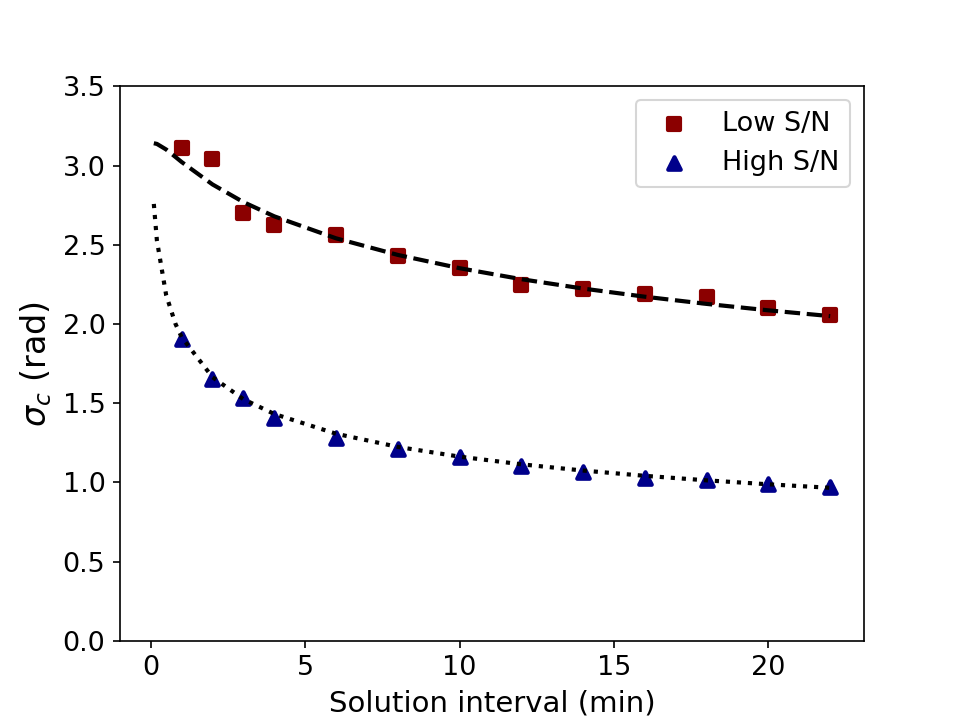}
  \caption{Phasediff-scores ($\sigma_{c}$) as a function of solution interval ($\delta_{t}$) for a high S/N calibrator (S/N = 1143) and a lower S/N calibrator (S/N = 122). The S/N is computed as the ratio of peak intensity to background RMS noise in the calibrated images, following the calibration procedure described in Section \ref{sec:internationalcal}. The values of $\sigma_{c}$ are derived by varying the solution intervals used for phase calibration to correct for right- and left-handed polarisation differences, and then computing the circular standard deviation of the resulting solutions. The dashed and dotted lines indicate the corresponding values derived with Equation \ref{eq:circscore}.}
\label{fig:phasediff_test}
\end{figure}

By varying the solution intervals between 1~min and 20~min for different calibrator sources when performing phase calibrations to correct for the differences between right-handed and left-handed polarisations, we empirically derived, with the two LOFAR observations considered in this work, the following relation between the phasediff-scores ($\sigma_{c}$) and the corresponding solution intervals ($\delta_{t}\,[\text{min}]$):
\begin{equation}\label{eq:circscore}
 \sigma_{c}=\pi\sqrt{1-\exp{\left(-\frac{\varsigma}{\sqrt{\delta_{t}}}\right)}} \quad [\text{rad}],
\end{equation}
where $\varsigma$ is a unique constant value for each calibrator, which must be determined individually for each source and each observation. This equation captures that increasing the solution interval reduces the variance in the phase solutions, whereas shorter intervals lead to higher variance. The value for $\varsigma$ can be understood as a measure related to the S/N of the source, as it indicates the extent to which the solution interval can be narrowed to achieve an acceptable phase solution variance on the calibration corrections. Figure \ref{fig:phasediff_test} shows that the equation provides a good fit to the measured circular standard deviation from phase calibration corrections for both low and high S/N calibrator sources. Notably, a root mean square error of 0.96 for the low S/N case and 0.99 for the high S/N case highlights the reliability of the fit. We find similar fits for other calibrator sources in our field and for both observations corresponding to a wilder and calmer ionosphere. In Appendix \ref{sec:circstat}, we relate Equation \ref{eq:circscore} to the theory behind circular statistics.

\begin{figure*}
 \centering
\includegraphics[width=0.42\linewidth]{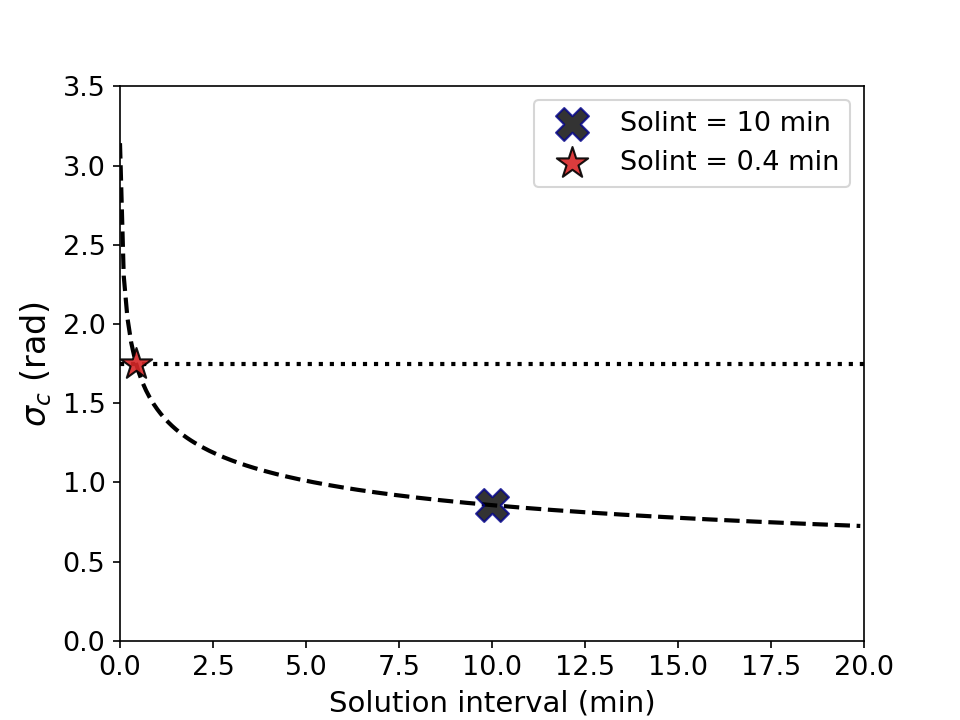}
\includegraphics[width=0.42\linewidth]{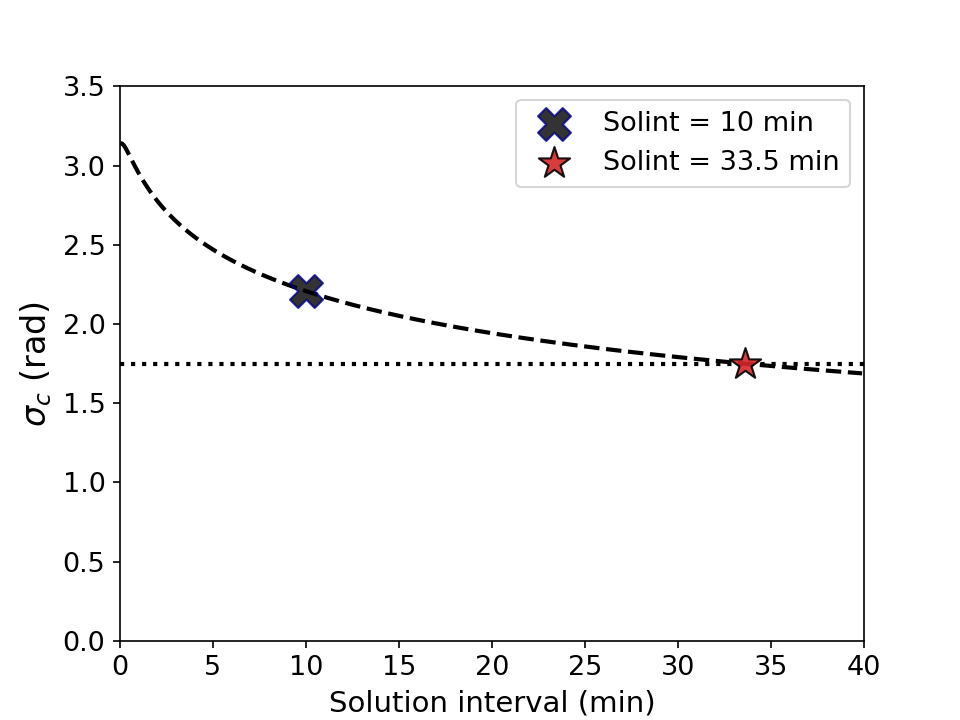}
  \caption{Phasediff-scores ($\sigma_{c}$) as a function of solution interval ($\delta_{t}$) for a high S/N calibrator source (\textit{left panel}) and a low S/N calibrator source (\textit{right panel}). The black cross is the phasediff-score calculated with the phase solutions for $\delta_t=10$~min, the dashed curve corresponds to the fit from Equation \ref{eq:circscore}, and the red star is the best fitted solution interval corresponding to $\sigma_c=1.75$, according to Equation \ref{eq:solint}.}
\label{fig:phasediff}
\end{figure*}

We derive $\varsigma$ for each source by evaluating $\sigma_{c}=\hat{\sigma}_{c}$ at a fixed solution interval of $\hat{\delta_{t}}=$10~min. This value corresponds to the same solution interval to select calibrator sources with phasediff-scores (see Section \ref{sec:selection}). By experimenting with self-calibrations of calibrator sources with different solution intervals, we found that $\sigma_{c}=1.75$~rad corresponds to a value of $\delta_{t}$ that serves as an effective metric for calibration stability (see Section \ref{sec:internationalcal}). This gives us enough information to derive from Equation \ref{eq:circscore} an expression for solution intervals by inserting $\hat{\delta}_{t}=10$~min and using the fact that $\varsigma$ is constant for each calibrator, such that
\begin{eqnarray}\label{eq:solint}
\delta_{t} & = & \hat{\delta_t} \left(\frac{\ln{\left(1-\left(\frac{\hat{\sigma}_{c}}{\pi}\right)^{2}\right)}}{\ln{\left(1-\left(\frac{\sigma_{c}}{\pi}\right)^{2}\right)}}\right)^{2} \quad [\text{min}] \nonumber \\ 
& \approx & 72.5 \cdot \left(\ln{\left(1-\left(\frac{\hat{\sigma}_{c}}{\pi}\right)^{2}\right)}\right)^{2} \quad [\text{min}].
\end{eqnarray}
This equation allows us to directly use the phase calibration solutions, for correcting for right-and left-handed polarisation on 10-min solution intervals, from the calibrator selection (see Section \ref{sec:selection}) and obtain a reliable estimate for the optimal solution interval for other types of calibration, as outlined in Section \ref{sec:internationalcal}. Figure \ref{fig:phasediff} illustrates for two calibrator sources how the solution interval from Equation \ref{eq:solint} can vary for different sources.

\subsection{Image inspection with neural networks}\label{sec:earlystopping}

By utilising metrics that combine commonly used image metrics, such as the RMS background noise, the dynamic range, and by visually assessing images and solution inspection plots, \cite{ye2024} and \cite{dejong2024} found that many of the calibrator sources required fewer or in some cases more self-calibration cycles than the often selected 10 or 12 cycles. In certain instances, self-calibration even began to diverge after just 5 or 6 cycles.
While the RMS and dynamic range provide useful information, they are not sufficient to fully determine if self-calibration has converged. For some calibrators, the RMS may for instance slightly increase during self-calibration when solving for amplitudes. In some cases, this is due to unstable diverging behaviour when a weaker calibrator absorbs signal from another source in the field, while in other cases, these are valid corrections to correct for small amplitude offsets. Similarly, the dynamic range may converge to favourable values during self-calibration, yet subtle image artefacts can persist and remain visible to the human eye. \cite{dejong2024} also examined calibration solution stability, which indicates whether the calibration solutions converge across different cycles, by comparing the difference between two sets of calibration solutions of the two following self-calibration cycles. While this confirms that self-calibration has indeed stabilised, it does not necessarily indicate that the resulting calibration is of sufficient quality, especially when overly conservative parameters are used, such as too long solution intervals or large frequency smoothing kernels.

\begin{figure*}
\includegraphics[width=1\linewidth]{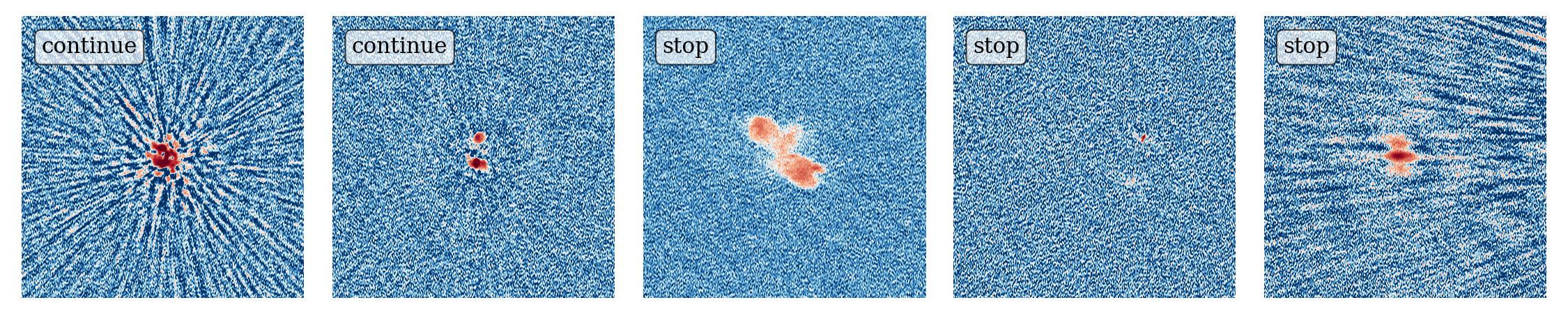}
\caption{Images of sources at various stages of self-calibration. \textit{Left:} Labelled \textit{continue}, as no self-calibration has been applied and significant artefacts are present. \textit{Middle left:} Labelled \textit{continue}, after several cycles of self-calibration, though noticeable artefacts remain. \textit{Middle:} Labelled \textit{stop} since self-calibration has successfully converged. \textit{Middle right:} Labelled \textit{stop} because there is not enough S/N to calibrate. \textit{Right:} Labelled \textit{stop} as artefacts from a nearby bright calibrator leak into the calibrator in the centre.}
\label{fig:nn_ex}
\end{figure*}

To address this, we remove the need for manual visual assessment by adopting a pre-trained DINOv2 neural network model \citep{oquab2023} with a custom classifier. In Appendix \ref{sec:nn_descript} we describe our model with our adjustments and methods to improve regularisation. We compiled a dataset of 3,000 binary-labelled self-calibration images from various observations. These observations include all calibrator sources used in \cite{dejong2024}, Bondi et al. (in prep.), and Escott et al. (in prep.), based on LOFAR data from the ELAIS-N1, Euclid Deep Field North (EDFN), and Bo\"otes Deep Field, respectively. The labelling was done by three expert radio astronomers to ensure that the model learns to mimic expert-level interpretation. The images include high S/N calibrator sources, weaker sources that were not selected for final DD calibration, and sources that are resolved out and thus correspond to images dominated by noise. We used 88\% of these images for training and left the remaining 12\% for validation. We further extend the training dataset by applying augmentations on our images, using random mirroring and rotations in multiples of 90~deg. This enhances the model's capability to train on images associated with more unique and complex extended calibrator sources.
The label \textit{continue} ($P=1$) is assigned to an image when there are significant artefacts and continued calibration is required. In all other cases, the images are labelled as \textit{stop} ($P=0$), which covers the following scenarios:
\begin{itemize}[leftmargin=*]
\item Self-calibration has converged.
\item The source is resolved out at high resolution, resulting in insufficient S/N for calibration.
\item There are strong artefacts in the image that are not originating from the central calibrator source.
\end{itemize}
Figure \ref{fig:nn_ex} presents examples of images with labels from our training dataset.

\begin{figure}
\centering
\includegraphics[width=0.8\linewidth]{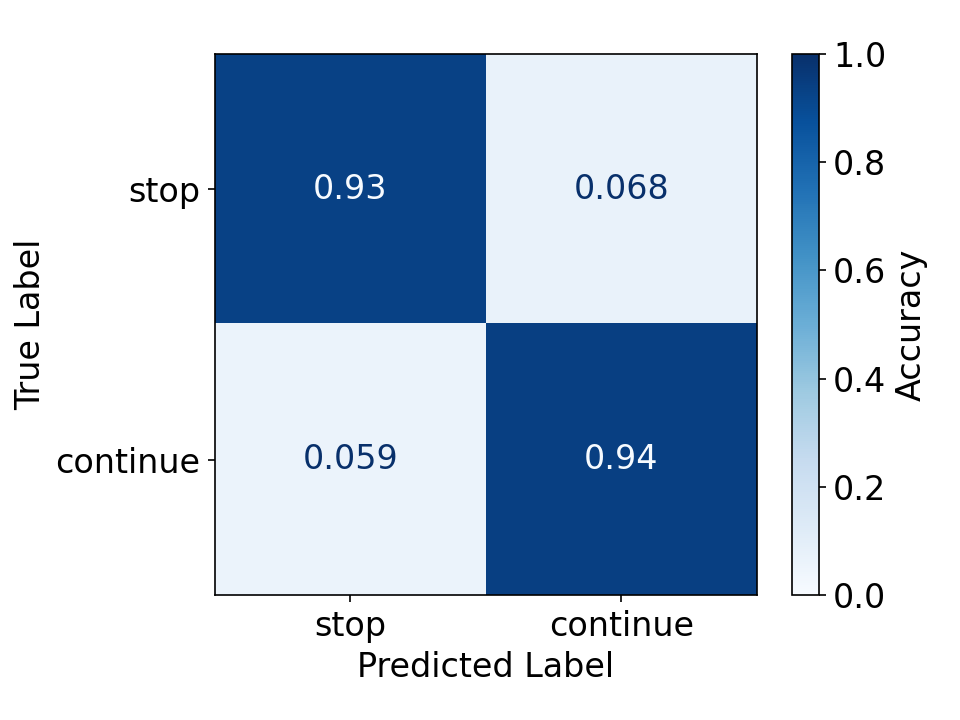}
\caption{Confusion matrix with the fraction of true positives, true negatives, false positives, and false negatives. The continue-continue and stop-stop labels indicate the accuracy of our model.}
\label{fig:cfmatrix}
\end{figure}

After training the model, we achieved an accuracy of 0.94 with the 360 validation images, which corresponds to the confusion matrix given in Figure \ref{fig:cfmatrix}. This indicates that the model maintains high precision in correctly labelling the sources in our validation dataset and according to our visual inspections.

\begin{figure*}
\includegraphics[width=1\linewidth]{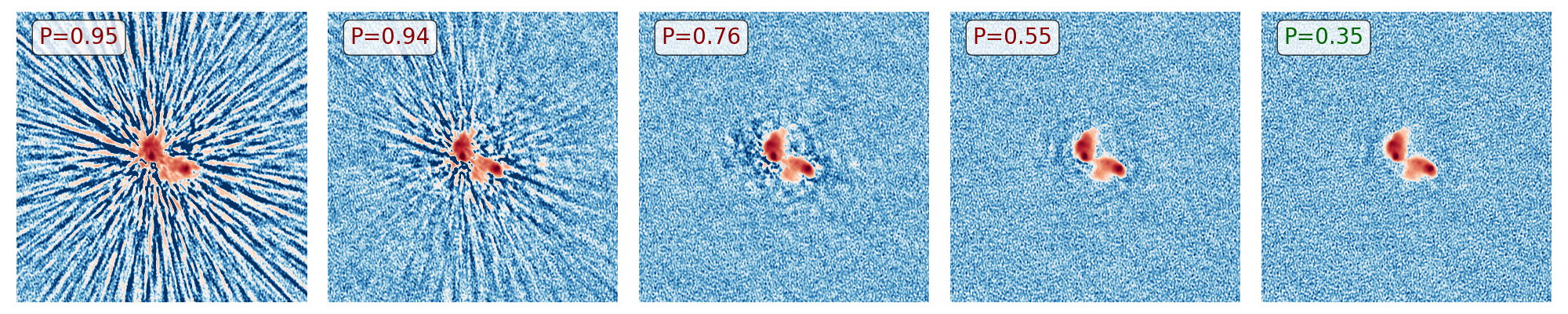}
\includegraphics[width=1\linewidth]{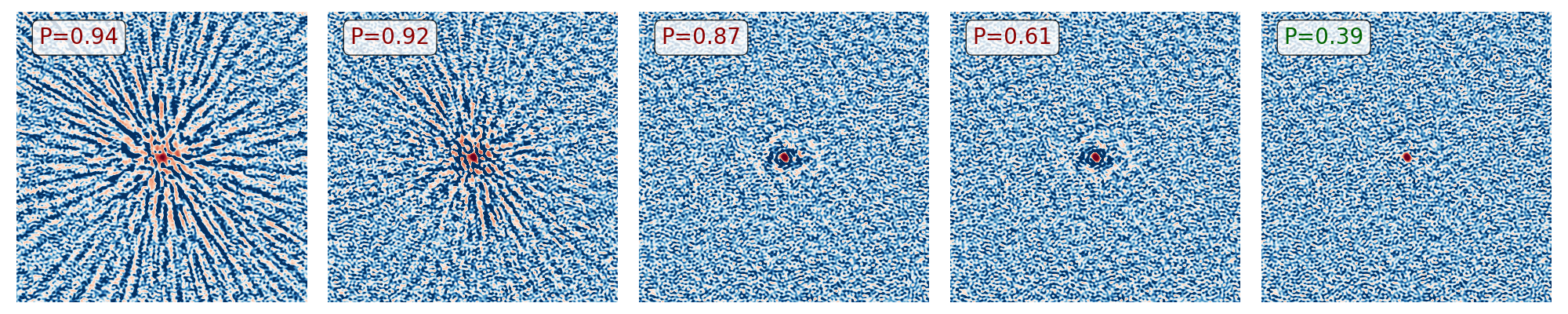}
\includegraphics[width=1\linewidth]{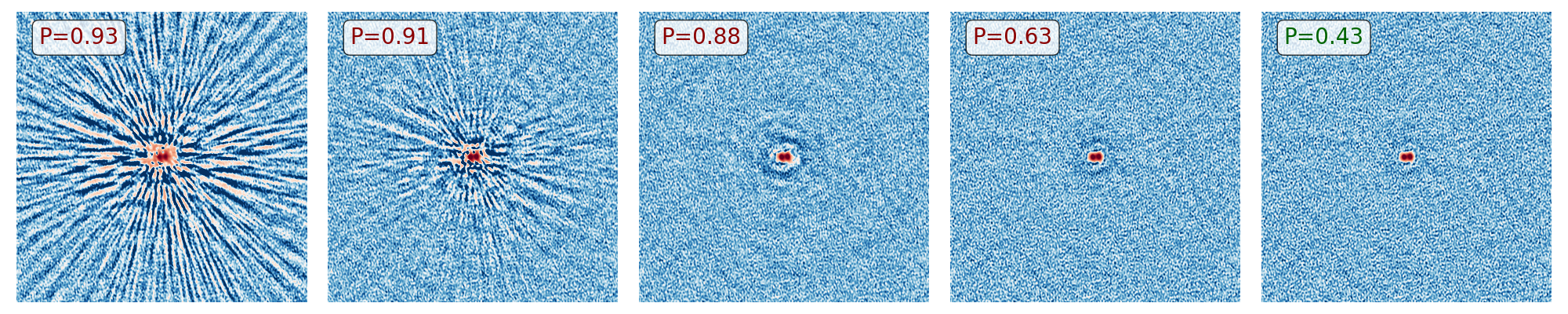}
\includegraphics[width=1\linewidth]{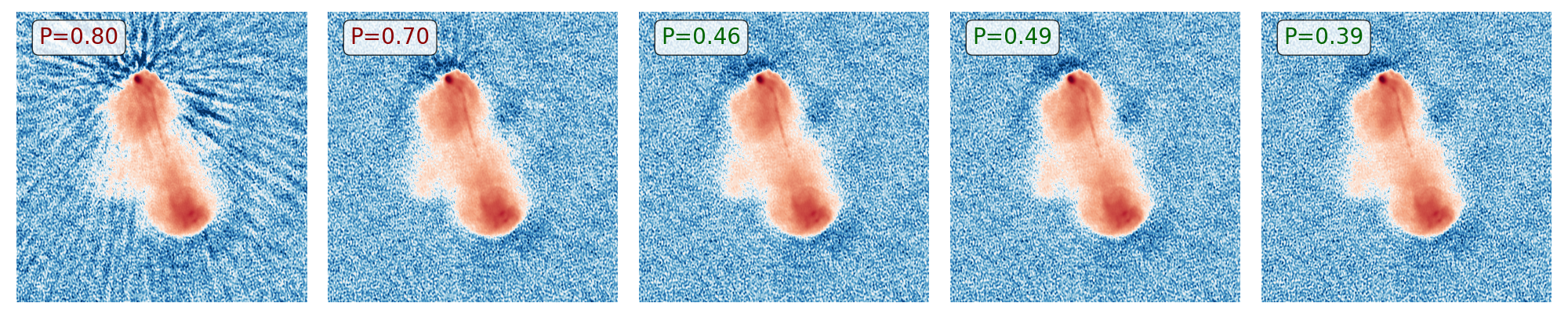}
\includegraphics[width=1\linewidth]{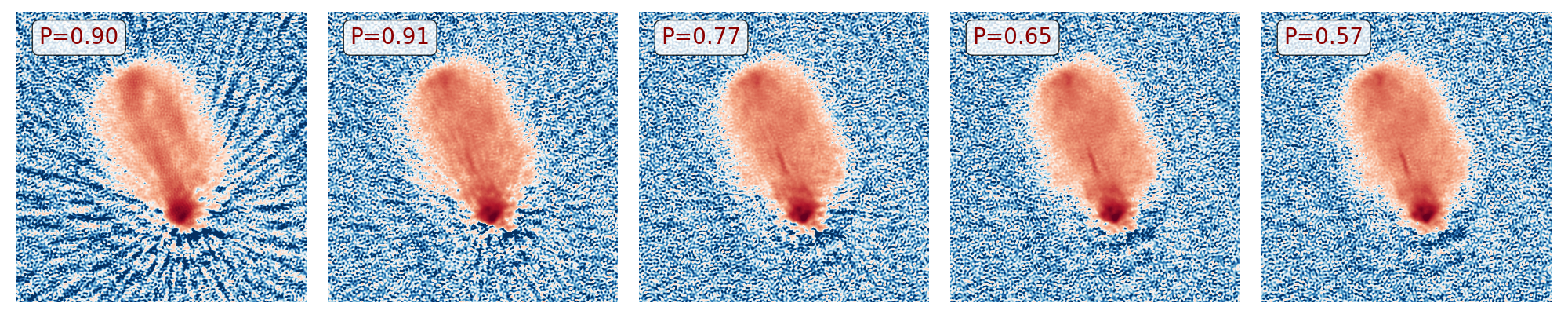}
\includegraphics[width=1\linewidth]{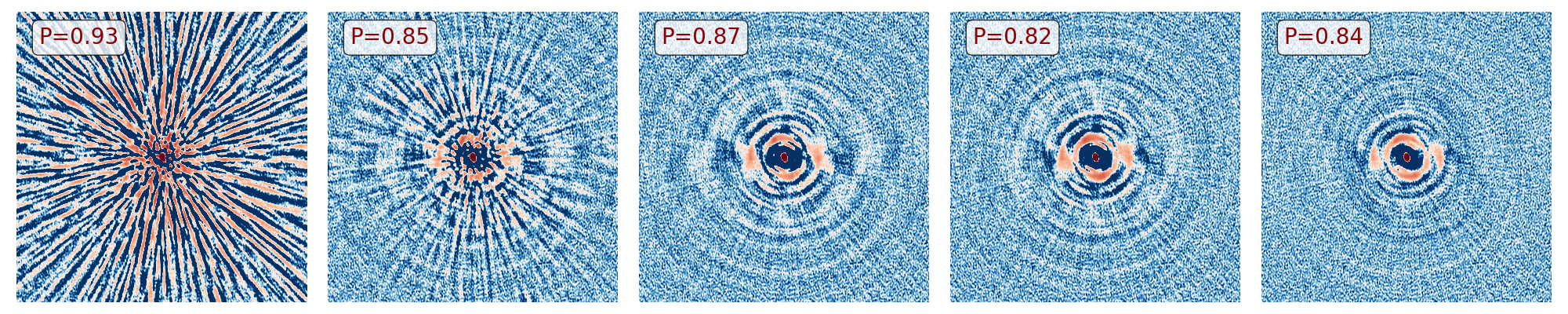}
\caption{Examples of self-calibration images with early-stopping scores from our neural network. The \textit{rows} represent different sources, while the \textit{columns} correspond to self-calibration cycles. $P<0.5$ values in green indicate successful convergence and let self-calibration stop, while $P\geq0.5$ in red suggests poor image quality, indicating that further self-calibration is required.}
\label{fig:pred_exs}
\end{figure*}

Figure \ref{fig:pred_exs} displays examples of label predictions for calibrators across various self-calibration cycles. The sources in the top three rows all achieve convergence by the 5th cycle, with $P$ values below 0.5. In each of these cases, we find that the model recognises subtle residual amplitude artefacts (4th column) that could still be corrected. In the fourth row, convergence is reached by the 3rd cycle, after which the label briefly increases, before reaching its lowest value in the 5th cycle. While the image shows notable improvement compared to the uncalibrated version, this example highlights the subjectivity of the labelling, as other astronomers might still find minor amplitude artefacts around the hotspots of this FRII source \citep{fanaroff1974} unsatisfactory.
In the fifth row, we observe a source gradually approaching convergence by the 5th cycle, though it has not yet met the early-stopping criteria for self-calibration. The final row illustrates a point source that fails to converge to an artefact-free image, although the model detects slight improvements compared to the initial uncalibrated image.

Since the model is trained purely to classify individual images, it does not account for amplitude drifts or information from prior self-calibration cycles. Therefore, we set our self-calibration stopping criteria on the requirements that the image is labelled with $P<0.5$, an RMS increase compared to the initial uncalibrated image of no more than 5\%, and an improved dynamic range compared to the uncalibrated image. We also require the solutions to have stabilised between cycles as well. This is determined by subtracting the current phase solutions from the previous one and calculating the circular standard deviation of the difference, which must be less than 0.1~rad. This threshold is based on the results from the self-calibration inspections from \cite{dejong2024}.
Our stopping criteria not only reduce the computational cost of self-calibration but also aid in identifying optimal calibration settings for our sources. This parameter optimisation allows us to fine-tune the automatic settings, as described in Section \ref{sec:internationalcal}.

\section{Refined strategy for facet-calibration}\label{sec:ddcal}

\begin{figure*}
\includegraphics[width=0.6\linewidth]{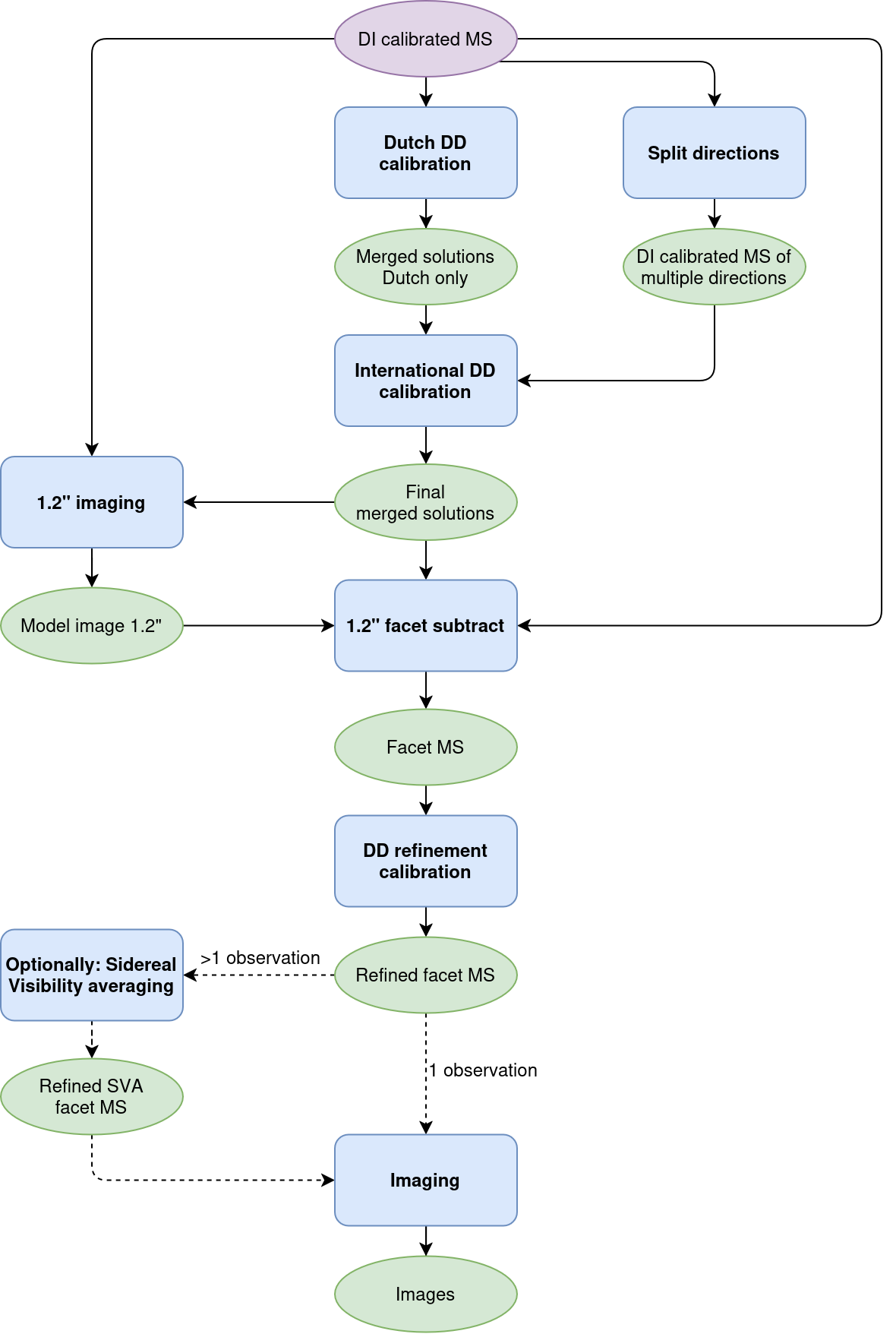}
  \caption{Flowchart corresponding to the full DD calibration discussed in Section \ref{sec:ddcal}. The workflow starts with DI-corrected datasets and ends with images. Purple ovals are input data, blue boxes are operations on the data, and green ovals are output data. When imaging more than one observation, it is possible to apply sidereal visibility averaging to combine datasets and boost imaging speed.}
\label{fig:ddworkflow}
\end{figure*}

In this section, we detail a DD calibration strategy up to the final imaging with the aim of enhancing image quality and reducing the computing costs compared to the work from \cite{dejong2024}. We follow the same data reduction strategy up to the primary in-field calibration as \cite{dejong2024}, where we only replace \texttt{prefactor} \citep{vanweeren2016, gasperin2019} by its successor, \texttt{LINC}, in the first calibration step. The outlined strategy corresponds to the flowchart from Figure \ref{fig:ddworkflow}. The workflow associated with this strategy is implemented with the Common Workflow Language \citep[CWL;][]{amstutz2016, crusoe2022} and will be part of the LOFAR VLBI pipeline \footnote{\url{https://git.astron.nl/RD/VLBI-cwl}}. Appendix \ref{sec:softwaredepen} provides an overview of the software dependencies for the calibration workflow.

\subsection{Dutch calibration}\label{sec:dutchcal}

Before the final imaging, \cite{dejong2024} generated separate datasets for each individual facet, enabling parallel imaging across facets. Sources outside each facet were subtracted using 1.2\arcsec~resolution model images with corresponding calibration solutions. However, it was found that the Dutch core and remote stations were poorly calibrated, where the issue became more pronounced when imaging at lower resolutions. This was evident both within individual facets and by the remaining artefacts from sources outside of these facets appearing across facet boundaries. 

The residual calibration artefacts likely stem from the limitations of the DD calibration strategy. This approach applies a high \textit{uv}-limit during self-calibration on an individual target calibrator, focusing on the calibration of the longer baselines. While effective for calibrating the international stations, this strategy is suboptimal for the Dutch LOFAR stations, which correspond to the short baselines and are therefore more sensitive to diffuse larger scale emission. Many sources within the field of view that are detected on short baselines become (partly) resolved out on the long baselines. When these sources are not included in the high-resolution sky model, they are more likely to degrade self-calibration performance for the Dutch stations than for the international stations. Furthermore, the international stations are more sensitive than the Dutch stations. As a result, sources that are high S/N for the international baselines have lower S/N on the Dutch baselines. Consequently, even though the high-resolution sky model may converge and improve calibration for the international stations, the Dutch stations may fail to fully converge, leading to unstable or inaccurate calibration solutions. These issues persist despite the use of the phase-up of the core to partially mitigate these effects \citep{morabito2022}. To improve the solutions from the Dutch stations, \cite{dejong2024} introduced an additional calibration refinement step for calibrating the Dutch stations alone after the 1.2\arcsec~resolution facet subtraction. While this improved image quality around the calibrator source of the facet, it did not eliminate residual calibration artefacts originating from neighbouring facets. Hence, it is essential that Dutch calibration solutions are properly corrected before any subtraction is performed.

To improve the calibration solutions for short baselines from the Dutch LOFAR stations, we introduce an additional DD calibration step for the core and remote stations before performing calibration with the international stations. We have available DD solutions from the \texttt{DDF}-pipeline\footnote{\url{https://github.com/mhardcastle/ddf-pipeline}} \citep{tasse2014, tasse2014b, smirnov2015}. However, we have already corrupted our data with the DI in-field calibration solutions for all LOFAR stations, which makes pre-applying Dutch DD solutions from the \texttt{DDF}-pipeline to the closest calibrator selected in Section \ref{sec:selection} no longer valid (see Section \ref{sec:futurecomp} for an alternative strategy to potentially mitigate this issue).

We opt instead for using a new joint-solve feature from \texttt{facetselfcal}, which performs self-calibration in multiple directions simultaneously using \texttt{DP3}. For this, we average our datasets to 16~sec and 195.36~kHz after removing the international stations. This reduces the data volume by a factor \textasciitilde300 and therefore reduces the computing resources required for a multi-directional joint-solve, while having enough time and frequency resolution to calibrate for the changing ionospheric effects.
Recent updates in \texttt{DP3} and \texttt{WSClean} have also enabled Stokes I-only data processing, which implies that 4 times less data has to be processed, saving RAM and computational time.
We select all calibrators that have a peak intensity above 85~mJy~beam$^{-1}$ in the 6\arcsec~ELAIS-N1 catalogue from \cite{sabater2021}, as we find this threshold to correspond to stable calibrators. This leaves us with 15 calibrators which we calibrate with the following strategy:
\begin{enumerate}[leftmargin=*]
\item To calibrate for fast varying phases for the most distant Dutch LOFAR stations, we first perform phase calibration of the Dutch remote stations. Given that brighter sources allow for shorter solution intervals, we use 16~sec for sources with peak intensities exceeding 300~mJy~beam$^{-1}$, and 32~sec for those below this threshold. We also use a frequency smoothness kernel of 20~MHz, which from experience has been shown to result in stable solutions at 6\arcsec. After calibration, we reset the solutions for the Dutch core stations by setting their amplitudes to 1 and phases to 0, ensuring that calibration corrections are applied only to the Dutch remote stations.
\item We then carry out phase calibration for slower phase variations from the Dutch core stations, using longer solution intervals of 64 seconds and a broader frequency smoothness kernel of 40~MHz. This approach works well for the Dutch core stations because their shorter baselines do not experience rapid phase variations as they primarily capture larger-scale structures.
\item After three cycles, we include calibration for the combination of phases and amplitudes. This applies longer solution intervals, compared to the previous phase calibration steps, since phases tend to vary on much shorter time scales compared to amplitudes. The solution intervals are 40~min for sources with peak intensities below 300~mJy~beam$^{-1}$, and 20~min for sources with peak intensities above this threshold. We use a frequency smoothness kernel of 10~MHz.
\end{enumerate}
All calibration steps are polarisation-independent, as polarisation corrections have already been applied through the \texttt{DDF}-pipeline full-Jones DI solutions and during a full-Jones in-field calibration step \citep[see Section 3.2.3;][]{dejong2024}.
The parameter settings mentioned above are optimised on an empirical basis for ELAIS-N1. For further optimisation, these settings may in the future be tied to S/N parameters, similar to the approach used for calibrating the full European array (as outlined in Sections \ref{sec:solints} and \ref{sec:internationalcal}).
This step corresponds to the `Dutch DD calibration' box and its merged output solutions in Figure \ref{fig:ddworkflow}.

\subsection{European calibration}\label{sec:internationalcal}

With the improved Dutch DD calibration solutions, we proceed to calibrate the international LOFAR stations. We first create datasets of each of the 24 selected calibrator sources (see Section \ref{sec:selection}), by phase-shifting to the centre of the source. We then average the datasets to 32~sec and 390.72~kHz. These averaging settings were shown by \cite{dejong2024} to be effective in suppressing signal from other high S/N sources in the field, while providing sufficient time and frequency resolution to correct fast phase variations, without introducing bandwidth or time smearing. Subsequently, we apply to each dataset the Dutch calibration solutions from the nearest of the Dutch calibrators from Section \ref{sec:dutchcal}. To further suppress the signal of other sources in the field of view and reduce the computational cost, we adopt, similar to previous works \citep[e.g.][]{moldon2015, morabito2022}, a phase-up of the Dutch core stations into a `superstation' (ST001).

Up to this stage, the Dutch stations have been DI and DD calibrated using a low-resolution sky model, which is adequate for 6\arcsec~imaging. However, incorporating the international stations adds longer baselines and requires a higher-resolution sky model, which can change the calibration solutions for the Dutch stations. As discussed in Section \ref{sec:dutchcal} and demonstrated by \cite{dejong2024}, phasing up the Dutch core and applying heavy averaging does not completely suppress contamination from sources in the field with high S/N on the shortest baselines that are not in the high-resolution sky model during self-calibration.
Solutions to mitigate this issue through source subtraction in visibility space exist, such as drawing boxes around calibrators and predicting and extracting all sources outside these boxes \citep[e.g.][]{vanweeren2021, dejong2022}. However, these methods are computationally too expensive for our large data volumes. Instead, we need to preserve during calibration the already obtained solutions for the Dutch stations and apply a high \textit{uv}-limit of 20~k$\lambda$, corresponding to angular scales smaller than 10.3\arcsec, to calibrate the international stations. This baseline limit corresponds to more compact source models during self-calibration and ensures the calibration focuses on the longest baselines.

\begin{figure*}
\includegraphics[width=0.48\linewidth]{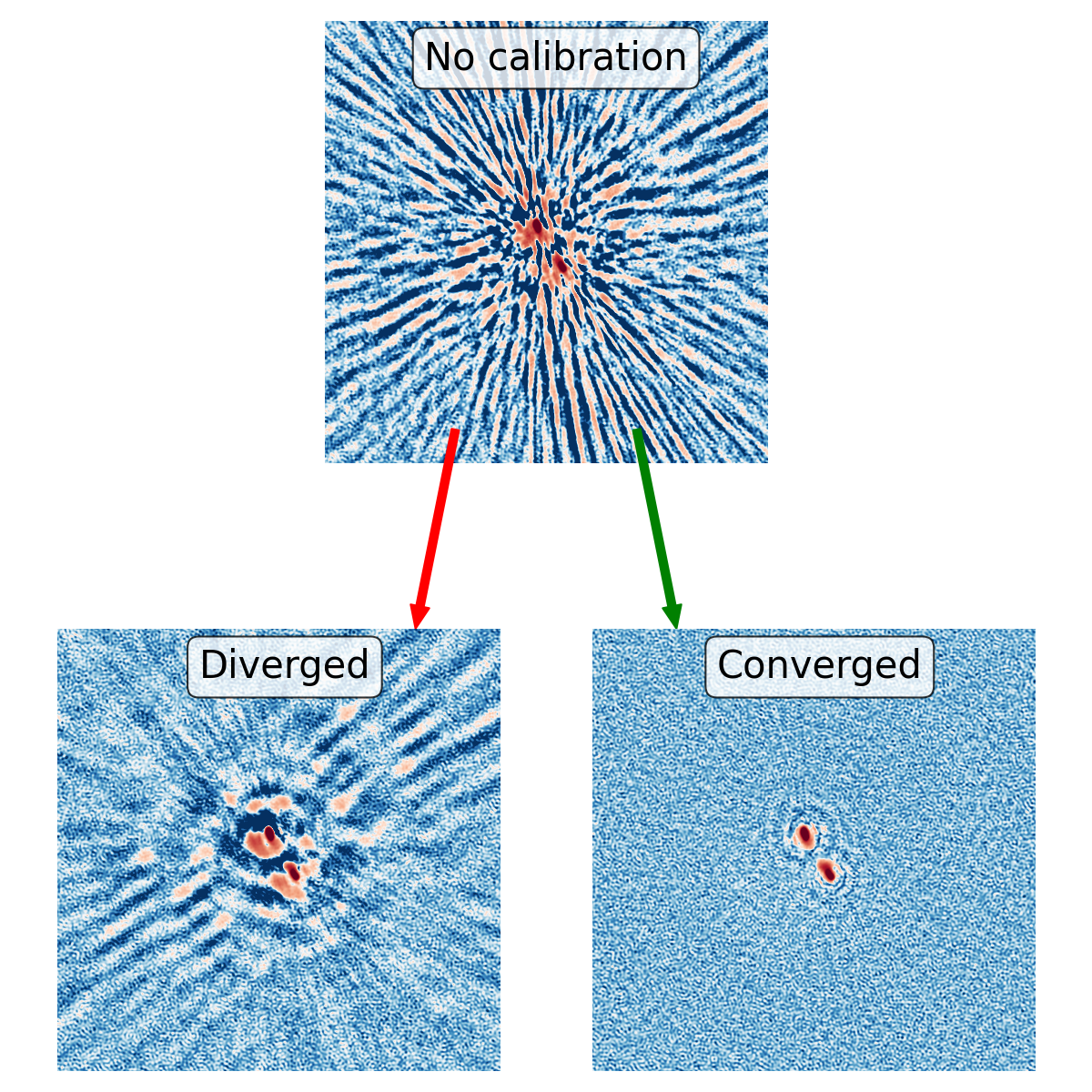}
\includegraphics[width=0.48\linewidth]{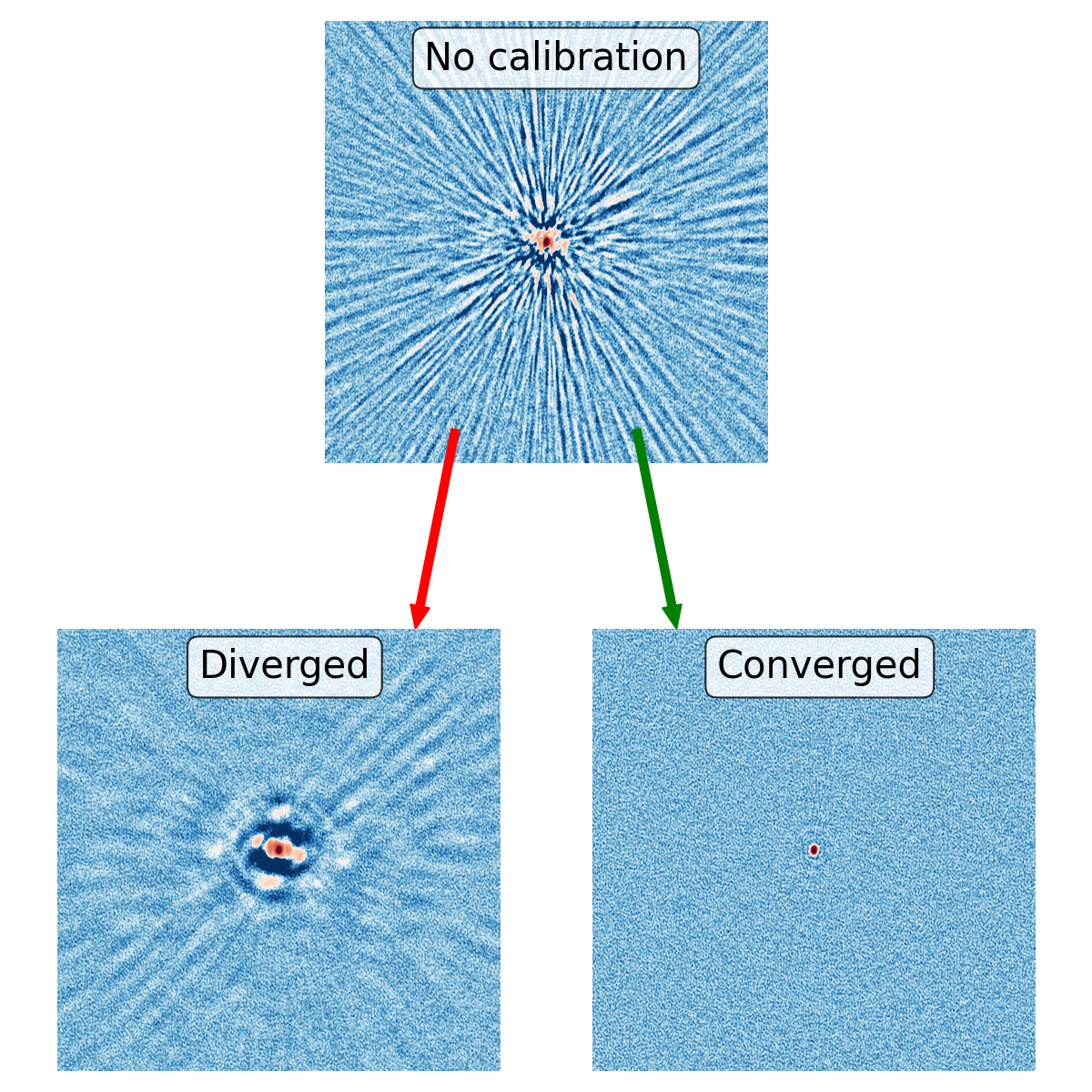}
  \caption{Self-calibration result for different calibration parameters. The top row displays two images of uncalibrated sources. Below, a left and right image illustrates the corresponding images after performing self-calibration on the uncalibrated sources, where the calibration either diverged or successfully converged based on different parameter settings. The difference in the settings between the diverging and converging images was dependent on changing the \textit{uv}-limit settings and by incorporating an additional round of polarisation-independent phase calibration for stations and resetting the Dutch core and a few additional remote stations differently, as we further outline in Section \ref{sec:internationalcal}.}
\label{fig:selfcal_stability}
\end{figure*}

\begin{figure*}
\centering
\fbox{\includegraphics[width=0.99\linewidth]{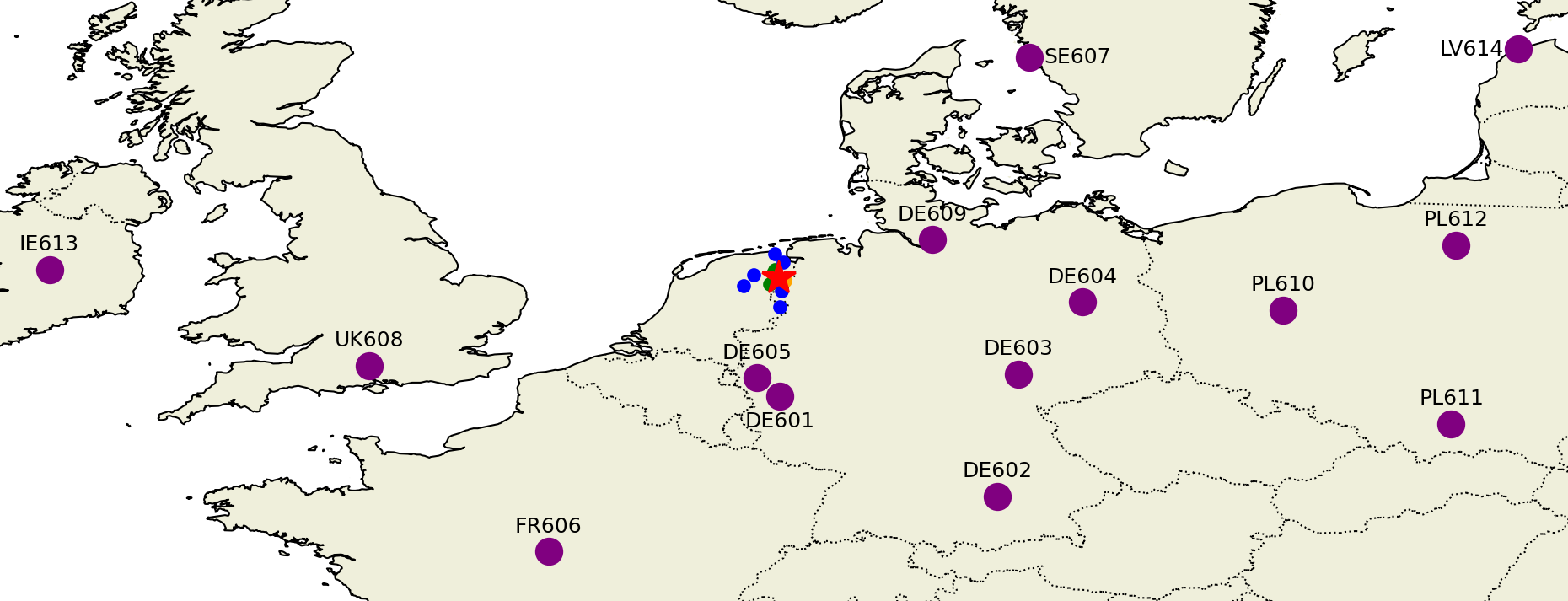}}
\vspace{1.5em}
\fbox{\includegraphics[width=0.5\linewidth]{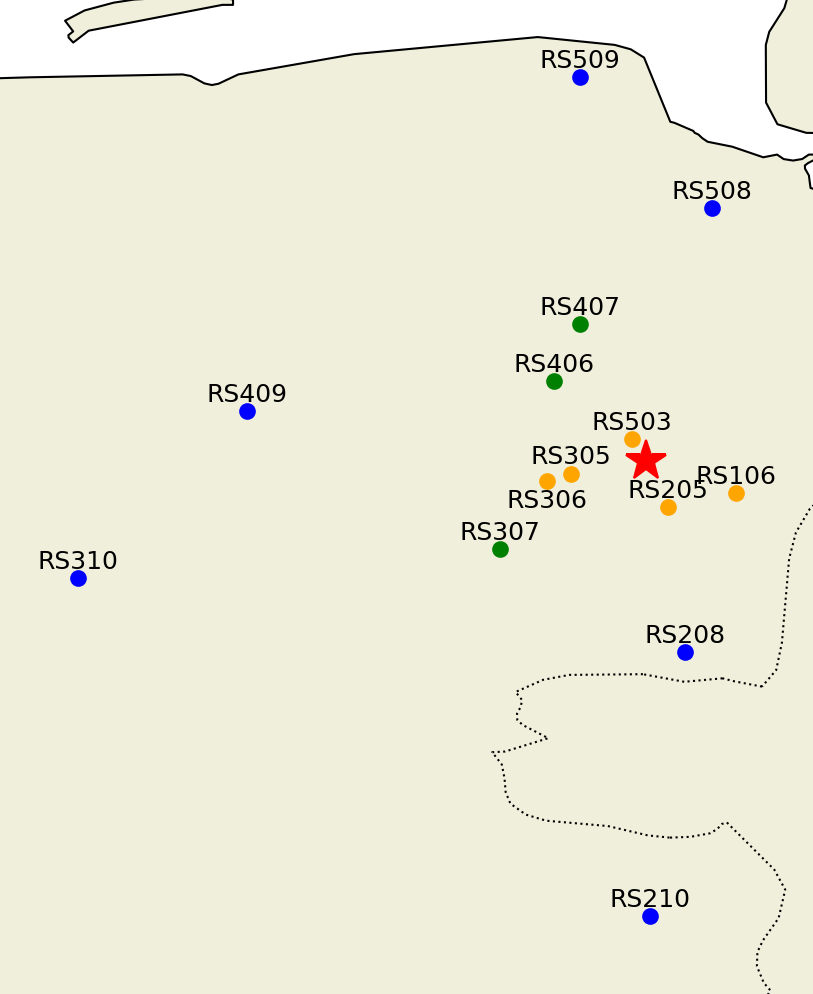}}
\caption{LOFAR station layout. The international LOFAR stations are highlighted in purple, the centre of the LOFAR core is indicated with a red star, the different groups of remote stations in blue, green, and orange. The \textit{upper panel} displays a zoom-out of the entire European array, while the \textit{lower panel} shows a zoom-in of the Dutch LOFAR stations.}
\label{fig:stationlayout}
\end{figure*}

We introduce multiple self-calibration steps, each targeting phase calibration for a different group of stations, with a final step also including amplitude calibration.
To calibrate specific station groups, we use the station reset option from \texttt{facetselfcal}, which resets the amplitude solutions to 1 and phases to 0 for a given station group. This provides a workaround for the limitation of the current \texttt{DP3} solvers, which cannot solve for different time intervals and frequency smoothness constraints for different stations. At each step, the Dutch core stations are reset, which means that their already obtained solutions (from Section \ref{sec:dutchcal}) are preserved. For calibrators with the lowest S/N, the remote stations are also reset at every step. In the case of higher S/N sources, we find improved results when (some of) the remote stations are allowed to be adjusted during calibration, which is because these sources are, due to their high S/N, less affected by contamination from other sources in the field. As a result, the $\delta_{t}$ parameter from Equation~\ref{eq:solint}, serving as a tracer for the S/N of a calibrator source, can be linked to which remote stations are reset in every step.
We also link $\delta_{t}$ to both the solution intervals and the frequency smoothing kernel used in each calibration step, as different station groups benefit from different solution interval lengths. In particular, the international stations, corresponding to the longest baselines, require corrections on shorter time and frequency scales compared to stations that are associated with most of the shorter baselines. To find the best combination of parameters and station group resets, we conduct a grid-search over different calibration parameter combinations and use the stopping criteria from Section \ref{sec:earlystopping} to distinguish between converging and diverging calibration behaviour (see Figure \ref{fig:selfcal_stability}). Based on our set of calibrators, we identify the following strategy as the most effective, as this corresponds to most of the calibrators having converged to images accepted by our neural network and with stable calibration solutions. Figure~\ref{fig:stationlayout} illustrates the different station groups referenced in this approach.

\begin{enumerate}[leftmargin=*]
\item We focus calibration first on the stations most distant from the Dutch core, by calibrating these using solution intervals with size $\sqrt{\delta_t/2}$ (using Equation \ref{eq:solint}) and frequency smoothness kernels of 8~MHz. The solution intervals have a minimal size equal to the time resolution of the datasets at 32~sec and a maximum size of 2~min. After this calibration step, we reset the solutions of all Dutch core and remote stations and German stations closest to the Dutch stations (DE601 and DE605) to preserve the already obtained calibration solutions of the Dutch stations and calibrate the closest German stations in the next step.
\item With the first phase solutions for the most distant LOFAR stations, we add another round of phase calibration, but now with larger solution intervals with size $\sqrt{\delta_t}$ and frequency smoothness kernels of 10~MHz. The solution intervals have a minimal size equal to the time resolution of the datasets at 32~sec and a maximum size of 3~min. After calibration, we reset the solutions for all Dutch core and remote stations.
\item The following step only continues for sources with solution intervals $\delta_t<3$~min, where we add another round of phase calibration for a specific (sub-)group of Dutch remote stations. Since this step involves solving for some of the Dutch stations, which correspond to shorter baselines, we use larger solution intervals and frequency smoothness kernels compared to the previous steps. The solution intervals are set to $2\sqrt{\delta_t}$ and the smoothness kernel to 15~MHz. For the brightest sources, with $\delta_t<0.3$~min, the best results were obtained by resetting only the solutions of the Dutch core stations, which implies that calibration solutions from all remote stations can be adjusted during self-calibration. For sources with $0.3\leq \delta_t<1$~min, we limit the adjustments for remote stations by resetting the solutions of all Dutch core stations and the five remote stations closest to the Dutch core (RS106, RS205, RS305, RS306, RS503). These are indicated in orange in the lower panel of Figure \ref{fig:stationlayout}. Lastly, for sources with $1\leq\delta_t<3$~min, we reset the same remote station, including a few more distant remote stations (RS307, RS406, RS407), which are indicated in green in the lower panel of Figure \ref{fig:stationlayout}.
\item Finally, after performing three cycles with only phase calibration, we also calibrate for both phases and amplitudes together but with larger solution intervals of $60\sqrt{\delta_{t}}$ with a minimum of 20~min, since phases vary on much shorter time scales than amplitudes. This step is entirely skipped when $20\delta_{t}>4$~hours. The smoothness kernels depends on the $\delta_{t}$, as we set this to 8~MHz if $\delta_{t}<1$~min, 10~MHz if $1\leq\delta_{t}<3$~min, and 12~MHz otherwise. We reset in this step the solutions from all Dutch core stations and the same specific sub-group of remote stations as for the previous step if $\delta_t\geq3$.
\end{enumerate}
All calibration steps are polarisation-independent, as we have already performed polarisation corrections during in-field calibration (see Section 3.2.3. in \cite{dejong2024}). As the Dutch stations were already well-calibrated, the shortest Dutch-only baselines had been corrected. This allows us to apply large \textit{uv}-limits and calibrate against a more compact sky model corresponding to the longest baselines. Combined with the station group resets, this approach helps to further suppress contamination from other sources in the field with high S/N on the shortest baselines (i.e., larger angular scales). Calibrator sources with higher S/N on the longest baselines can have larger \textit{uv}-limits while still achieving self-calibration convergence, whereas lower S/N calibrators require shorter remote station baselines to provide enough signal for reaching convergence. In combination with the calibration strategy outlined above, we found the following empirically derived relation to be effective in further improving our results:

\begin{equation}\label{eq:uvcut}
\text{\textit{uv}-limit} = 40 - 20\cdot\exp{\left(-\frac{1}{\delta_{t}}\right)} \quad [\text{k}\lambda].
\end{equation}
This scales the \textit{uv}-limit to values between 20 k$\lambda$ and 40 k$\lambda$, where the former corresponds to a largest angular scale of approximately 10.3\arcsec, while the latter excludes more remote-remote baselines and corresponds to a largest angular scale of around 5.2\arcsec.

For deciding on early-stopping during self-calibration, we utilise the neural network in combination with the other quality metric assessments discussed in Section \ref{sec:earlystopping}. We use between 3 and 20 self-calibration cycles, providing a large enough range to ensure self-calibration convergence for most calibrator sources. For deep imaging involving multiple observations, we do not need to perform self-calibration with all datasets combined, as this will be addressed during the final self-calibration step (see Section \ref{sec:selfcal2}). This allows in this intermediate calibration step for embarrassingly parallel processing over each calibrator source and each observation. The step discussed in this Section corresponds to the `International DD calibration' box and its merged output solutions in Figure \ref{fig:ddworkflow}.

\subsection{Facet subtraction for sub-arcsecond imaging}\label{sec:subtract}

To reduce the wall-clock time for 0.3\arcsec~wide-field imaging, we divide, similar to \cite{sweijen2022} and \cite{dejong2024}, the full dataset into smaller subsets through averaging over time and frequency, where each dataset corresponds to a single facet. This enables parallel imaging of all facets on smaller datasets. To avoid calibration errors from leaking into the facet that is being imaged, it is important to subtract all sources that do not correspond to this facet. This subtraction is performed by predicting visibilities using 1.2\arcsec~model images, which can be generated 16 times faster than 0.3\arcsec~resolution models. This is because the visibilities for 1.2\arcsec~imaging can be averaged by a factor of 4 in both time and frequency compared to those required for 0.3\arcsec~imaging, without introducing additional bandwidth or time smearing effects, and we image 16 times fewer pixels. \cite{dejong2024} associated the 1.2\arcsec~imaging, prediction, and subtraction step as part of the final imaging, as it utilises imaging software (\texttt{WSClean}) for prediction of sources within the field of view and it was the final part in the pre-processing stage before the actual imaging at sub-arcsecond resolution. This step accounted for 76\% of the total imaging costs, which represents 62\% of the overall computational costs, including all calibration steps. The remaining 24\% of the imaging costs were attributed to imaging the facets using all observations combined, as discussed in Section \ref{sec:imaging}. This implies that reducing the computational cost at this stage can significantly lower the overall computational cost of the total data processing.

To lower the computational costs of the source subtraction step before imaging, we introduce an improved method for creating datasets for each facet. Instead of copying the full datasets for each facet at the full data resolution (1~sec and 12~kHz), masking a single facet, and predicting and subtracting all sources outside the facet, we perform model data predictions on datasets averaged to 4~sec and 48~kHz and predict each facet only once. The time and frequency resolutions are the same as used for making the 1.2\arcsec-resolution model images, ensuring that there is little to no loss in prediction accuracy compared to predicting at 1~sec and 12~kHz, while lowering the computational costs by a factor of 16. The prediction and subtraction strategy from \cite{sweijen2022} and \cite{dejong2024} also led to redundant predictions of the same sky, which with our new strategy is only done once per facet, providing another large computational cost reduction. To generate model visibilities for subtraction when creating datasets for each facet, we sum all the model visibilities from all the facets excluding the one corresponding to the dataset, and use nearest-neighbour interpolation to restore the resolution to 1~sec and 12~kHz. We then subtract the summed interpolated model visibilities from the full visibility dataset. Finally, we phase-shift and average to the facet centre before applying the corresponding scalar calibration corrections from the DD-calibration and performing a beam correction. The total computational costs reduction compared to the old method and taking into account I/O overhead, brings us to a computational cost reduction for this step of a factor of 10 (see Section \ref{sec:cpucost}). The subtraction steps correspond to the `1.2\arcsec facet subtract' in Figure \ref{fig:ddworkflow}.

\begin{figure}
\includegraphics[width=1\linewidth]{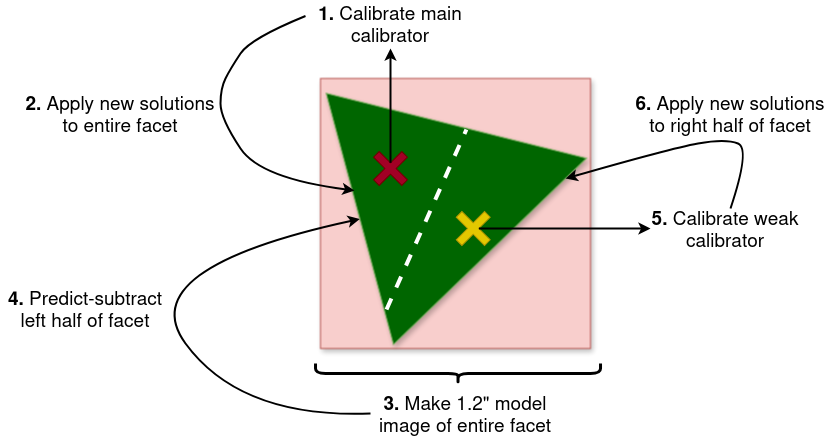}
  \caption{Additional facet calibration refinement for facets with weaker calibrators with low phasediff-scores ($2.3\leq\sigma_{c}<2.6$~rad). This calibration involves several steps. First, the main facet calibrator is calibrated using all observations combined. The derived solutions are then applied to the facet, and a 1.2\arcsec~model image of the entire facet is created. With this model, the left half of the facet is predicted and subtracted, where both halves are defined with a Voronoi tessellation. This allows for the calibration of a secondary, weaker calibrator, after which the new solutions are applied to the remaining half of the facet. Finally, both halves of the facet are imaged.}
\label{fig:refinement}
\end{figure}

\subsection{Final calibration refinement}\label{sec:selfcal2}

Since the sources with low phasediff-scores ($\sigma_{c}<2.3$~rad), indicating high S/N at the longest baselines, are now contained in separate datasets corresponding to individual facets, we mitigate the interfering effect that high S/N calibrators have on each other when they are not corrected for DDEs and are contained in the same dataset. This enables us to apply a final joint-calibration step to improve calibration accuracy for all baselines and all observations together when performing deep imaging.



For the final calibration step, we employ the same approach as we described in Section \ref{sec:internationalcal}. This means that we phase-shift back to the centre of the calibrator sources corresponding to each facet before averaging the datasets back to 32~sec and 390.72~kHz and calibrate each source according to the strategy outlined in Section \ref{sec:internationalcal}. The only difference is that we apply the third and fourth calibration steps to all calibrator sources without resetting the calibration solutions for the Dutch stations, thereby enabling joint calibration of all LOFAR stations. While this final calibration step may be considered optional when calibrating and imaging a single LOFAR observation, it becomes essential for deep imaging that combines multiple observations, since it marks the first stage where data from all observations are aligned. This ensures that small astrometric offsets between datasets are corrected and data is calibrated to the same model data during self-calibration.

Finally, there is an optional step to include calibration of the weaker `secondary' DD-calibrators, which were defined in Section \ref{sec:selection} as sources with phasediff-scores of $2.3\leq\sigma_{c}<2.6$~rad. These sources may not have high enough S/N to have their own facet during the previous stages of the calibration, but they can still suffer from DDEs that can be corrected through an additional refinement step. The need for this step depends largely on the severity of the ionospheric conditions, which vary per observation. However, because the facet datasets are smaller at this stage, these refinement steps are significantly less computationally expensive than earlier calibration and prediction steps. In this refinement, we first create a 1.2\arcsec~model image of the entire facet and apply a Voronoi tessellation, using the positions of the main facet calibrator (which has already been corrected and solutions have been applied to the data) and the weaker calibrators. The area corresponding to the main facet calibrator is predicted and subtracted, by averaging by a factor of four in time and frequency (on top of the already applied averaging from Section \ref{sec:subtract}) and using the method discussed in Section \ref{sec:subtract}. This is followed by a phase-shift to the weaker calibrator before further calibration, using the same method as for the main facet calibrators. This process splits the facet into two (or more) sub-facets, where the weaker calibrator benefits from the pre-applied solutions of the primary calibrator, allowing for further refinement of local DDEs, and faster final imaging of the smaller datasets. Figure \ref{fig:refinement} illustrates this process, beginning with the calibration of the primary calibrator and showing the subsequent steps leading to the final calibrated data. In the case of multiple weaker calibrators, this can be easily extended to multiple facet splits, where all weaker calibrators have their own local solutions.
The entire calibration refinement step, as described in this subsection, corresponds to the `DD refinement calibration' box and its dataset output in Figure \ref{fig:ddworkflow}.

\subsection{Imaging}\label{sec:imaging}

Once the calibrated datasets for each facet have been obtained, we proceed with imaging with \texttt{WSClean} \citep{wsclean}. Following the approach of \cite{dejong2024} and as employed in earlier imaging stages, we use the \texttt{wgridder} module \citep{arras2021, ye2022}. For the imaging of each facet, we apply Briggs weighting with a robust parameter of $-1.5$ \citep{briggs1995}, a minimum \textit{uv}-limit of 80$\lambda$, and pixel sizes of 0.1\arcsec, 0.2\arcsec, and 0.4\arcsec. These are combined with Gaussian tapers of 0.3\arcsec, 0.6\arcsec, and 1.2\arcsec, respectively. For efficient deconvolution, we use automatic masking, multi-scale cleaning \citep{cornwell2008}, and an RMS box size set to 50 times the synthesised beam \citep{offringa2017}. A final correction for the primary beam attenuation is also applied.

\section{Data volume compression}\label{sec:datavolcomp}

The initial unaveraged and uncompressed input LOFAR data is about 16~TB. These datasets are in the LOFAR Long-Term Archive (LTA) in most cases compressed to 4~TB, using Dynamical Statistical Compression \citep[Dysco;][]{offringa2016}. Despite the already applied lossy compression, the data volumes remain large and pose a significant bottleneck for processing the LOFAR datasets. In particular, for sub-arcsecond wide-field imaging, where we want to image the longest baselines at high time and frequency resolutions, without loss of image quality. These data resolution constraints lead to high computational costs. We highlight three key methods in the context of reducing radio interferometric data volumes: 
\begin{itemize}[leftmargin=*]
\item \textit{Lossy compression}: Approximating the data by reducing precision or modelling it with representative distributions, while introducing only minimal or acceptable errors into the data \citep{offringa2016, dodson2024}.
\item \textit{Baseline-dependent averaging (BDA)}: Making use of the fact that shorter baselines can be averaged more in time and frequency compared to longer baselines \citep[e.g.][]{cotton1986, cotton2009, wijnholds2018}.
\item \textit{Sidereal visibility averaging (SVA)}: Averaging visibilities at similar baseline coordinates in the specific case of imaging multiple observations taken over multiple sidereal days \citep{dejong2025}.
\end{itemize}
Lossless compression methods also exist, but are not considered here, as they are partly integrated within lossy compression software and, on their own, are significantly less effective on noise-dominated data compared to lossy compression \citep[e.g.][]{peter2017}.
In the following subsections, we briefly discuss the above-mentioned methods in the context of high-resolution imaging with LOFAR.

\subsection{Lossy compression}\label{sec:lossy}

The current standard method for lossy compression of LOFAR data is Dysco \citep{offringa2016}. This uses quantisation to store data with fewer bits and normalisation by grouping visibilities that have a similar distribution. This implies that noisy low S/N data with a Gaussian distribution may be more compressible than high S/N data in the sense that, for a given compression factor, the absolute increase in noise is smaller in low S/N data compared to high S/N data. In the LTA, data are compressed by default with a factor of 4, with visibility data stored at 10 bits and their associated weights at 12 bits per 4 correlations. These settings generally lead to a negligible loss of information \citep[e.g.][]{chege2024}. For high-resolution imaging with all LOFAR stations, these settings can be considered conservative, since data is in this case kept at high time and frequency resolutions and are therefore typically more noise-dominated than the more averaged datasets from the Dutch array alone.
With the default Dysco settings, visibility weights are already compressed by a factor of 10 compared to uncompressed data \citep{offringa2016}, since Dysco automatically stores only one polarisation for the visibility weights and uses a 12-bit compression, leaving limited potential for further compression relative to the visibilities (only in the order of a few per cent). Visibilities, on the other hand, are stored as complex values with all polarisations (not in Stokes-I). As a result, reducing the number of bits to store visibilities offers the greatest potential for data volume reduction beyond the default Dysco settings.

\begin{figure}
\includegraphics[width=1\linewidth]{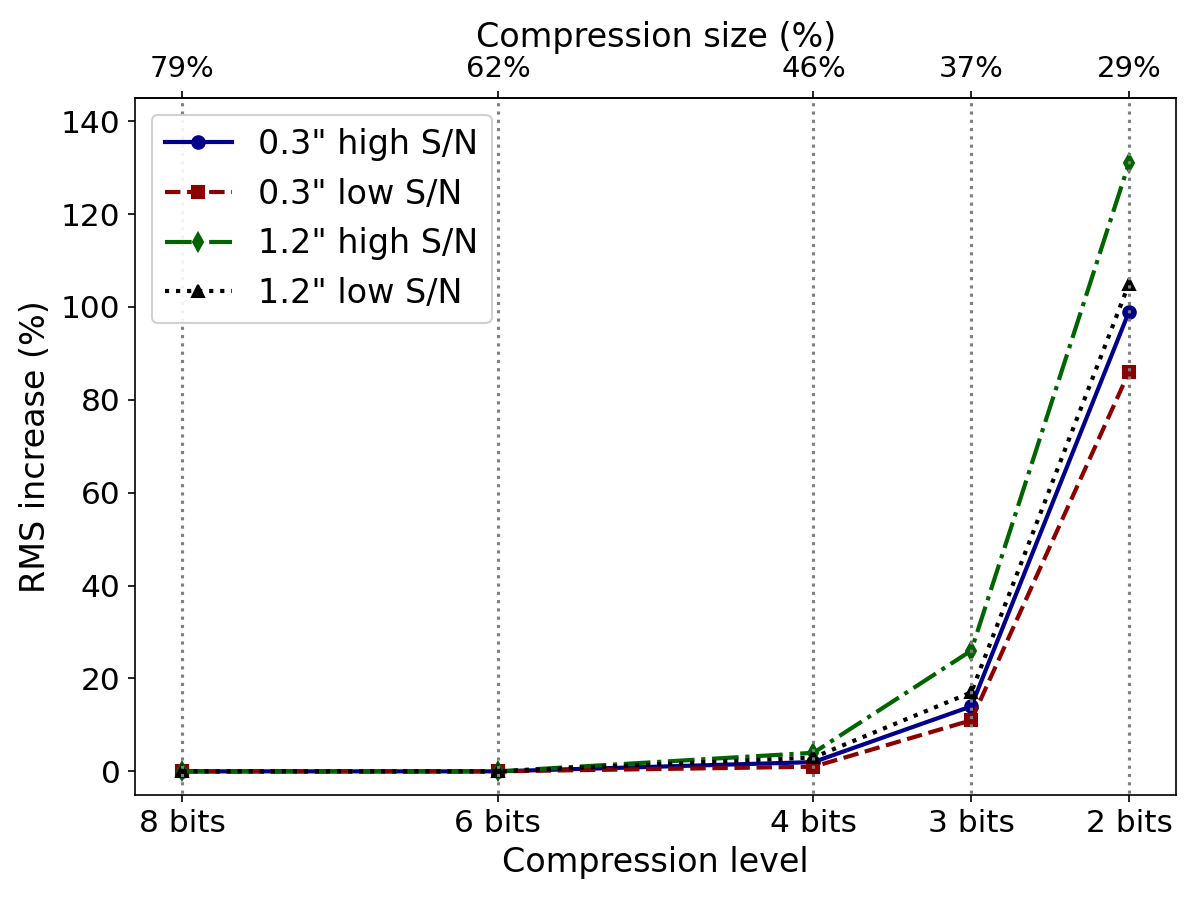}
  \caption{Comparison of RMS noise increase in image space as a function of the number of bits the visibilities are stored and compression size for two different facets, imaged at two different resolutions. The 1.2\arcsec~images are averaged four times more in time and frequency compared to the 0.3\arcsec~images. The compressed size is compared with Dysco's default settings (10-bit visibility storage and 12-bit weight storage).}
\label{fig:datavol}
\end{figure}

To determine the extent to which we can compress our data, we selected two facets from the ELAIS-N1 data after the 1.2\arcsec~subtraction, where one of the facets corresponds to a high S/N calibrator and the other to a low S/N calibrator. For varying compression levels expressed in the number of bits, we show in Figure \ref{fig:datavol}, the RMS increase in the images as a function of compression level, together with the data volume compressed size compared to the default 10-bit visibility and 12-bit weight compression with Dysco.
This figure indicates that with 6-bit compression, the RMS background noise remains unchanged, and only an increase in the order of a few per cent is observed for 4-bit compression, with more pronounced noise increases at higher compression levels, reaching about 100\% at 2-bit compression. We also find that the image quality at 1.2\arcsec~imaging is more affected at higher bit rates to RMS increases compared to the 0.3\arcsec~resolution imaging. This is because the data is more averaged over time and frequency, making the data less noise-dominated, and therefore less compressible. We verified that the residual images at these bit rates, obtained after subtracting the image corresponding to the original 10-bit stored data, remain purely noise-dominated and show no unusual artefacts. Additionally, we confirmed that the peak intensities remain unaffected. Our results demonstrate that for the ELAIS-N1 dataset, and likely for many other LOFAR pointings, the visibility data can be compressed to 6-bits, resulting in a 40\% data volume reduction compared to the default settings or to a 85\% data volume reduction compared to uncompressed data. In cases where data volume presents a significant bottleneck, such as large-scale data processing with limited storage, 4-bit compression may in some cases also be viable. We recommend including a step in the pipeline to find the appropriate compression for a given dataset.

\subsection{Baseline-dependent averaging}\label{sec:bda}

As discussed in the previous subsection, lossy compression is an effective method to reduce data volumes. However, this compression method does not reduce the number of visibilities, which implies that although you save in storage space, you do not achieve gains in terms of processing speed. For this purpose we need to reduce the number of visibilities being processed without introducing bandwidth or time smearing or compromising image fidelity. This can be done by applying baseline-dependent averaging (BDA).

BDA enables different time and frequency averaging for different baselines. This leverages the fact that shorter baselines experience slower phase variations over time and frequency, allowing their visibilities to be averaged more compared to longer baselines that are more sensitive to phase changes \citep[e.g.][]{cotton1986,wijnholds2018}. Too much BDA can degrade resolution, increase the severity of already existing calibration artefacts, and introduce smearing artefacts for sources more distant from the pointing centre. Based on the equations presented in \cite{bridle1999}, the average peak intensity loss factor ($\tau$) over an observation of a source at the celestial pole is taken as
\begin{equation}\label{eq:bridle}
\tau = 1- 1.22\cdot10^{-9}\left(\frac{\theta\cdot\Delta t}{\theta_{\text{res}}}\right)^{2},
\end{equation}
where $\Delta t$ is the required time resolution, $\theta$ the angular distance from the phase centre in radians, and $\theta_{\text{res}}$ the angular resolution in radians. 
Given that the angular resolution depends on both frequency ($\nu\,[\text{Hz}]$), baseline length ($B\,[\text{m}]$), and the speed of light ($c\,[\text{m/s}]$) through 
\begin{equation*}
\theta_{\text{res}}=\frac{c}{\nu\cdot B},
\end{equation*}
we can invert Equation \ref{eq:bridle} to determine the maximum time resolution over which a baseline can be averaged. This gives
\begin{equation}\label{eq:timebda}
\Delta t = 8.7 \cdot 10^{12} \frac{\sqrt{1-\tau}}{\nu \cdot B \cdot \theta},
\end{equation}
Hence, for a given observation targeting a specific image size and allowing a fixed peak intensity loss tolerance, the time resolution per baseline scales with this formula inversely with baseline length as $\Delta t \propto B^{-1}$.
BDA is therefore particularly effective for LOFAR observations utilising the full European array, which encompasses a wide range of baseline lengths, from approximately 0.13~km up to 2,000~km.

Although Equation~\ref{eq:bridle} is derived as an average over an observation of a source at the celestial pole, and is therefore not a precise measure of instantaneous time smearing, it provides an estimate for time smearing as a function of baseline length, as expressed in Equation~\ref{eq:timebda}. In reality, the behaviour of time smearing is more complex, depending on the observing geometry and the rate at which each baseline moves through the \textit{uv}-plane. In addition, smearing in the resulting images also depends on imaging parameters, particularly the Briggs weighting, which applies varying weights to different baselines. Nevertheless, for the purposes of this work, we adopt the simplified baseline-length-based scaling as a first-order estimate. This approximation is sufficient for estimating the impact of time smearing introduced by BDA on our data, as we demonstrate below.

\begin{figure}
\includegraphics[width=1\linewidth]{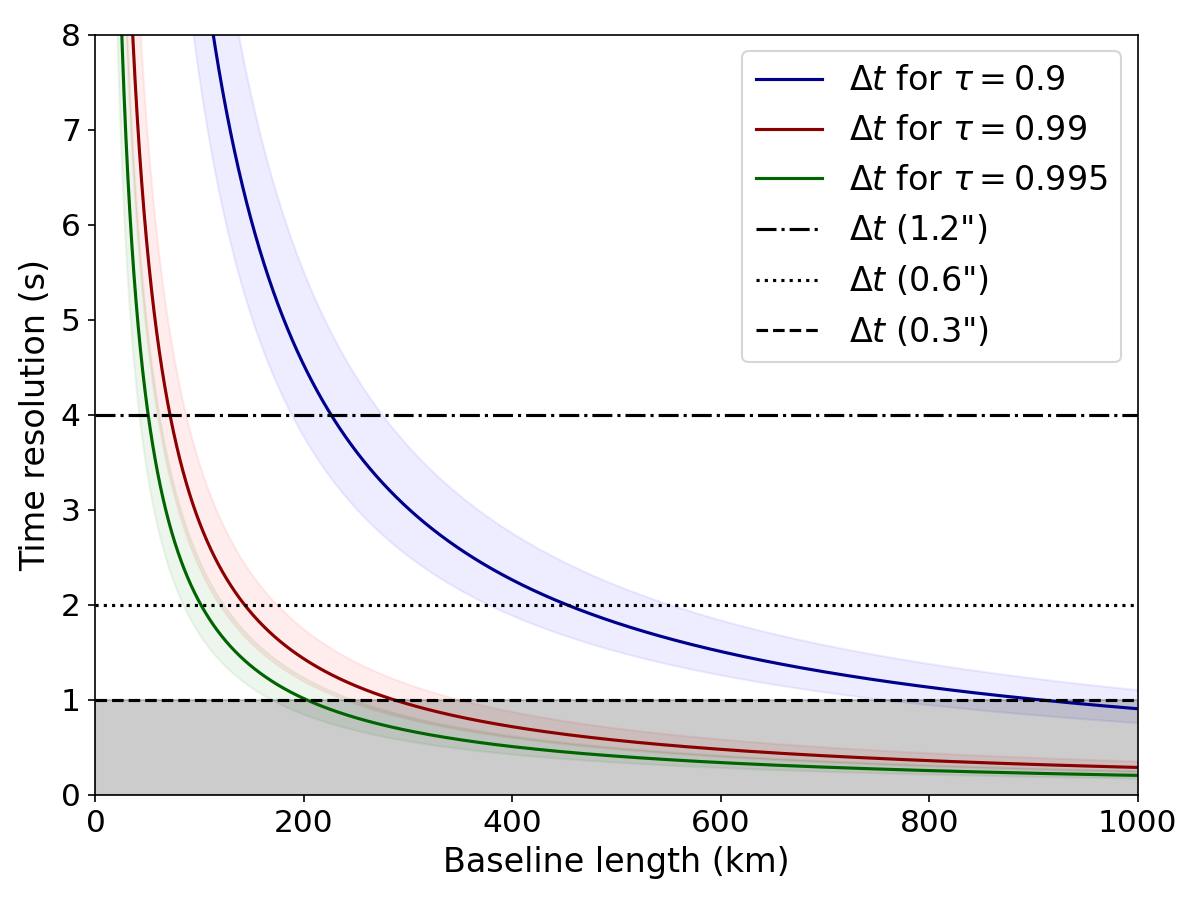}
  \caption{Time resolution as a function of baseline length for different peak intensity loss factors ($\tau$), as estimated by Equation \ref{eq:timebda}, together with the time resolutions for 1.2\arcsec, 0.6\arcsec, and 0.3\arcsec~imaging. The gray shaded area highlights the time resolution limit given by the 1~sec data resolution of LOFAR data stored in the LTA. The shaded areas around the green, red, and blue lines correspond to the 118 to 168~MHz range of our data (see Table \ref{table:metadata}).}
\label{fig:bdatimebase}
\end{figure}

\begin{figure}
\includegraphics[width=1\linewidth]{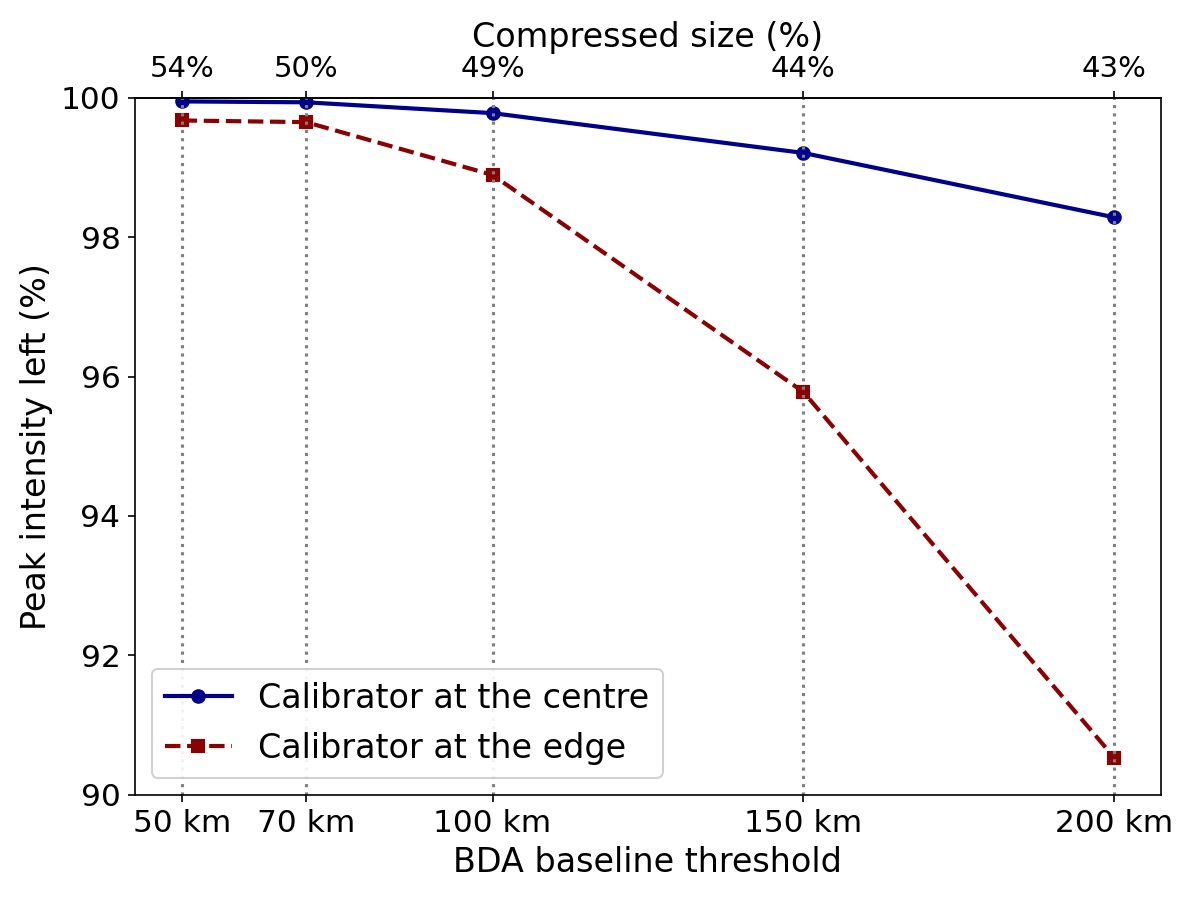}
  \caption{Comparison of peak intensity left in the 1.2\arcsec~resolution wide-field image when imaging with BDA as a function of the baseline length threshold beyond which BDA is not applied. We also added the visibility compression size. To compare sources at different distances from the pointing centre, we used two high S/N calibrator sources located near the centre of the field and the other near the edge of the field. The maximum BDA time averaging is in all cases set at 32~sec, which corresponds to the smallest solution interval.}
\label{fig:peakintensbda}
\end{figure}

\begin{figure}
\centering
\includegraphics[width=0.99\linewidth]{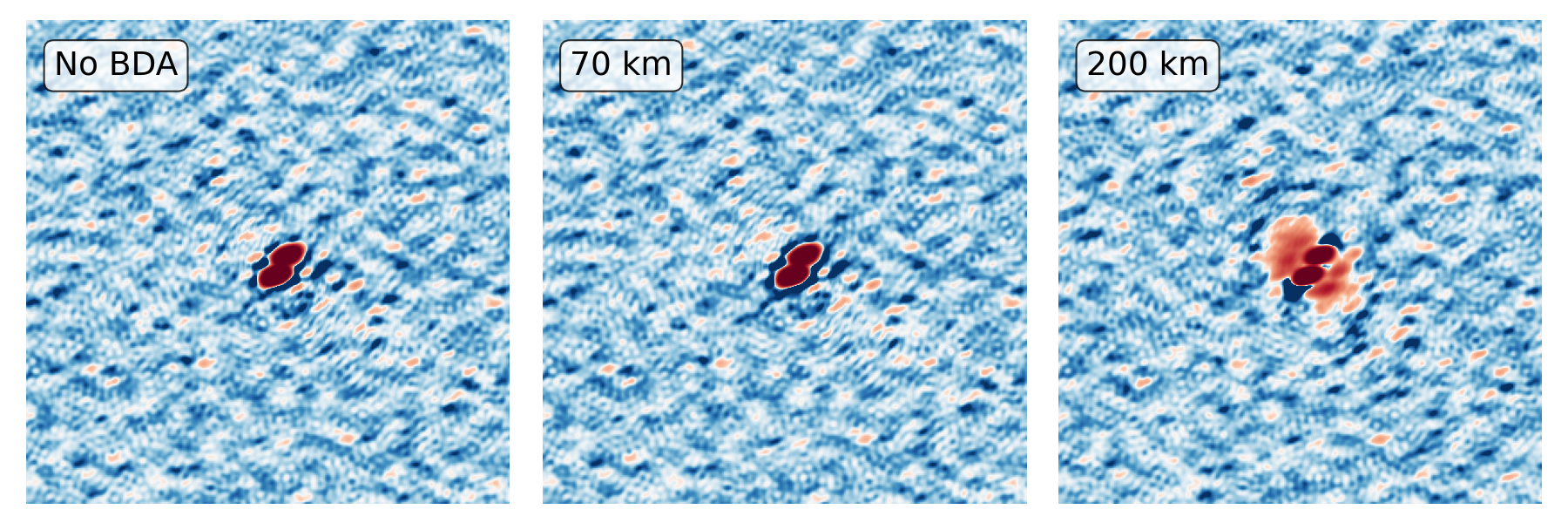}
\includegraphics[width=0.99\linewidth]{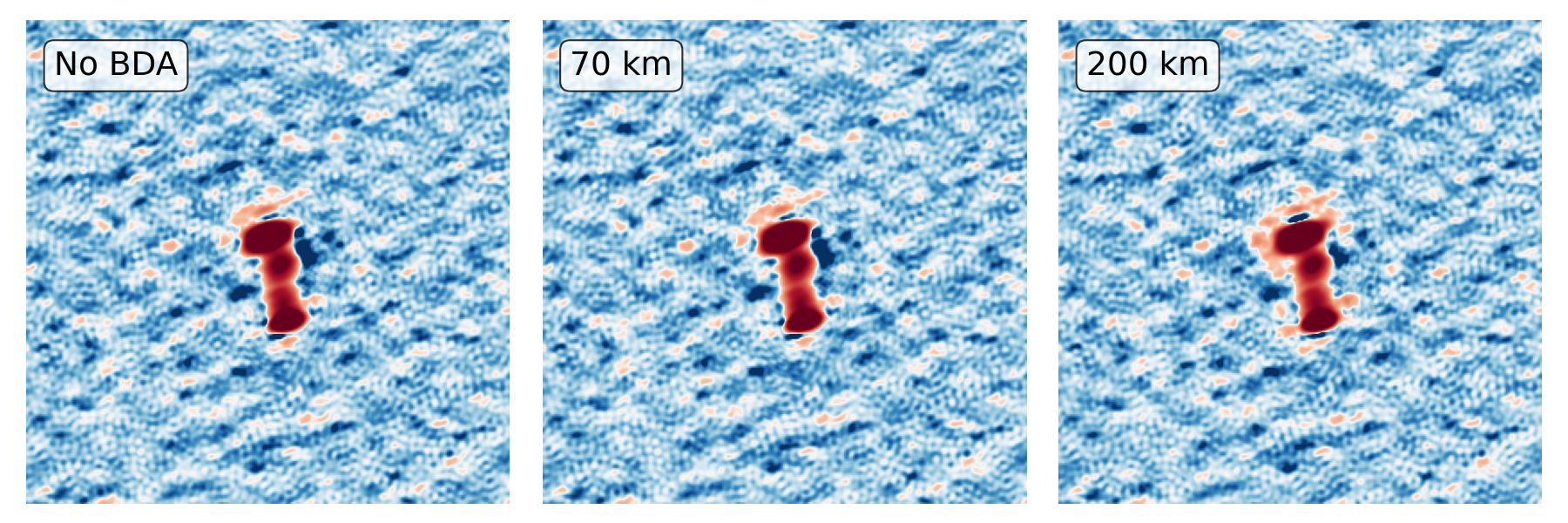}
  \caption{Image quality comparison of two sources (\textit{rows}) imaged at 1.2\arcsec~resolution using datasets without BDA (\textit{left}), with BDA applied up to 70~km baselines (\textit{middle}), and up to 200~km baselines (\textit{right}), where the data was originally averaged to 4~sec. These are the same sources as in Figure \ref{fig:peakintensbda}, with the top row showing the source near the edge of the field and the bottom row the one near the centre.}
\label{fig:bdasources}
\end{figure}

We tested different BDA settings with \texttt{DP3}. However, we are currently restricted to BDA in the time domain, since \texttt{WSClean} does not currently support imaging of datasets with BDA in the frequency domain. This limits us to optimise two parameters from the \texttt{DP3} implementation:
\begin{itemize}[leftmargin=*]
\item The maximum time averaging with BDA to limit the averaging at the shortest baselines.
\item The baseline length beyond which BDA is not applied. The BDA time averaging factor is then equal to this value divided by the baseline length, rounded to the nearest multiple of the time resolution.
\end{itemize}
For the first parameter, we use 32~sec, as this matches the shortest solution interval employed in our DD calibration and thus represents a safe choice (see Section \ref{sec:internationalcal}). To estimate the BDA baseline threshold, we use Equation~\ref{eq:timebda} to find the time resolution as a function of baseline length for various peak loss factors ($\tau$) with an angular distance of 1.25~deg, as shown in Figure~\ref{fig:bdatimebase}. This figure compares these time-baseline curves with the time resolutions for the three main target resolutions (1.2\arcsec, 0.6\arcsec, and 0.3\arcsec). The points of intersection represent the baseline lengths beyond which BDA should not be applied. We find for instance for 1.2\arcsec~imaging at 4~sec that a threshold of below 70~km is, according to Equation \ref{eq:timebda}, appropriate to limit the peak intensity loss to less than 1\%, while for imaging at 0.3\arcsec~at 1~sec we find a BDA baseline threshold of 280~km for the same peak intensity loss. To compare these estimates with real data, we show in Figure \ref{fig:peakintensbda} the peak intensity change in image space for two high S/N calibrator sources, when performing 1.2\arcsec~imaging without BDA and imaging with BDA for different baseline length thresholds. Images of these sources with different BDA settings are shown in Figure \ref{fig:bdasources}. These results show that smearing effects for a high S/N source near the edge of the field increase beyond thresholds of 70~km, which agrees with having a value of $\tau\approx0.995$ in Figure \ref{fig:bdatimebase}. With a baseline threshold at 70~km, the number of visibilities are reduced by about a factor of 2, which speeds up imaging by about a factor of 1.5.

\subsection{Sidereal visibility averaging}\label{sec:sva}

Deep imaging with multiple observations is typically carried out by repeatedly gridding and degridding the visibilities from all observations combined. This approach introduces a considerable computational overhead. By leveraging the fact that baseline tracks repeat every sidereal day, sidereal visibility averaging (SVA) can be used to reduce the number of visibilities by averaging those with comparable baseline coordinates \citep{dejong2025}. A similar method was proposed by \cite{owen2008} but the recent revisit for LOFAR by \cite{dejong2025}, using fully calibrated datasets, also takes into account effects like nutation, precession, and aberration that can significantly affect the \textit{uv} coordinates of the longest baselines. When applied to multiple calibrated LOFAR observations, SVA leads to substantial reductions in computational cost during the final imaging stage (see Section \ref{sec:imaging}). This reduction does not scale linearly with the number of observations, as the SVA algorithm introduces its own computational overhead.

\section{Results}\label{sec:results}

In this work, we have implemented enhancements in calibration quality and computational efficiency for processing LOFAR data for sub-arcsecond wide-field imaging. To demonstrate their effectiveness, we have processed two datasets. In this section, we highlight the resulting improvements in terms of calibration and computing costs.

\subsection{Image quality}\label{sec:calqualimpr}

The calibration strategy has been updated by addressing challenges highlighted in previous work \citep{dejong2024}, and incorporating the solution interval metric and neural network described in Section \ref{sec:automated}. This has led to changes in image quality, as outlined in this subsection.

\subsubsection{Artefact suppression at facet boundaries}\label{sec:facetbound}

\begin{figure*}
\includegraphics[width=0.9\linewidth]{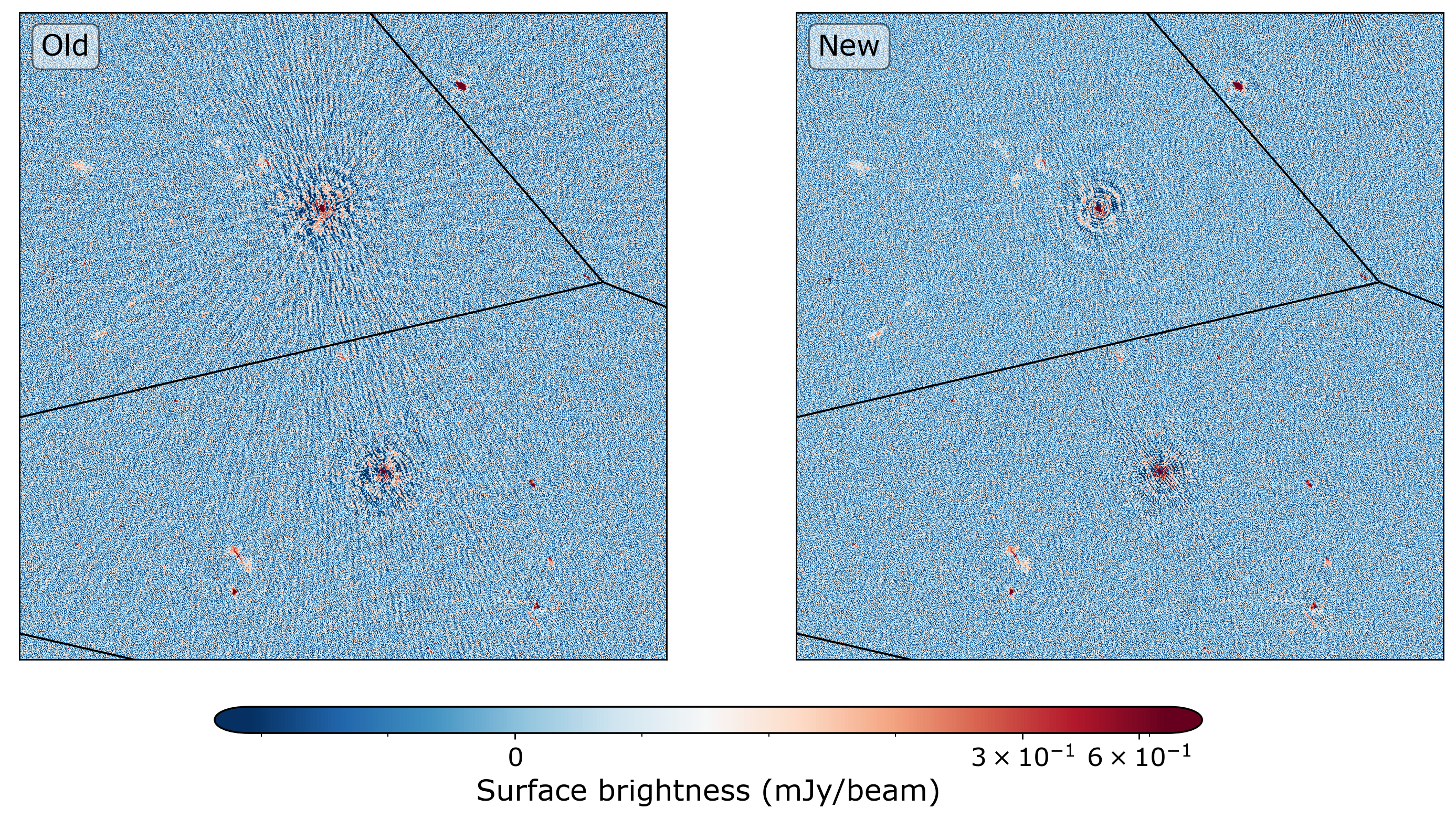}
\caption{Improvements in direction-dependent (DD) calibration for 1.2\arcsec~imaging for the observation with the wildest ionosphere (\texttt{L833466}). The \textit{left} image displays the image quality using the calibration strategy from \protect\cite{dejong2024}, while the \textit{right} image shows the results employing the new calibration strategy, where we first apply Dutch-only DD calibration before calibrating the international stations. The black lines represent the facet boundaries.}
\label{fig:12improved}
\end{figure*}

In Section \ref{sec:dutchcal}, we introduced an additional step to perform DD calibration for the Dutch core and remote stations before proceeding with DD calibration for the international stations. This enhances the Dutch calibration solutions, which is in particular important for producing the 1.2\arcsec~model images required for subtracting sources outside each facet before imaging (see Section \ref{sec:subtract}). Figure \ref{fig:12improved} illustrates the improvements in the 1.2\arcsec~resolution image by comparing the imaging of a challenging region by using the old \citep[from][]{dejong2024} and the new (this work) calibration strategies, for the observation with the wildest ionosphere. In the left panel of this figure, DD artefacts from two neighbouring calibrators were previously leaking into each other's facets. It is clear that in the new image in the right panel, this effect has been mitigated, which enables performing an improved subtraction of the source signal at short (Dutch) baselines.

\subsubsection{Varying ionospheric conditions}

\begin{figure*}
\centering
\includegraphics[width=0.99\linewidth]{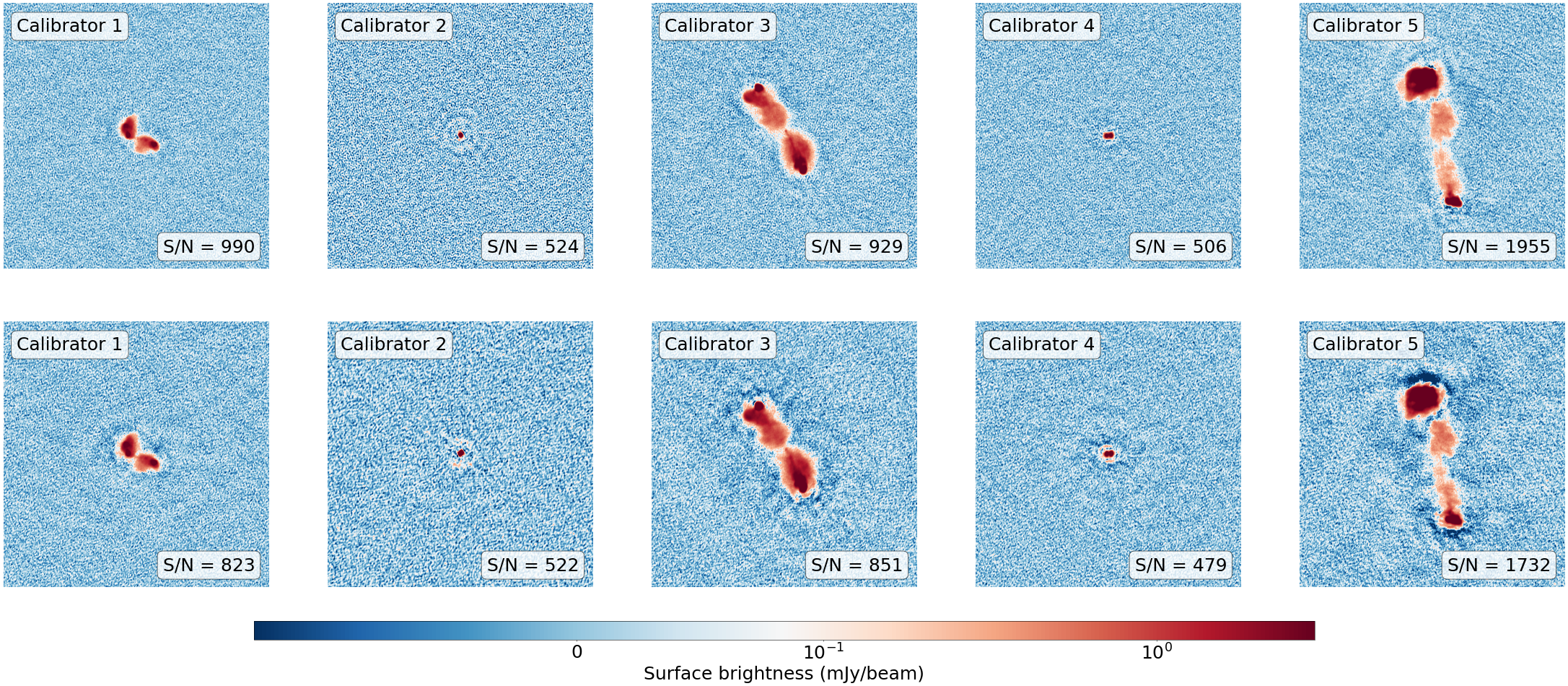}
  \caption{Comparison of self-calibration results for five different calibrator sources across the two observations selected in this work. The \textit{upper} panel shows results for observation \texttt{L686962}, which was conducted under relatively calm ionospheric conditions. The \textit{lower} panel corresponds to observation \texttt{L833466}, taken under more disturbed ionospheric conditions. The S/N reported in the image is calculated as the peak intensity divided by the local background noise, providing a measure of the relative brightness of each source.}
\label{fig:selfcal_comparison}
\end{figure*}

\begin{figure*}
\centering
\includegraphics[width=0.99\linewidth]{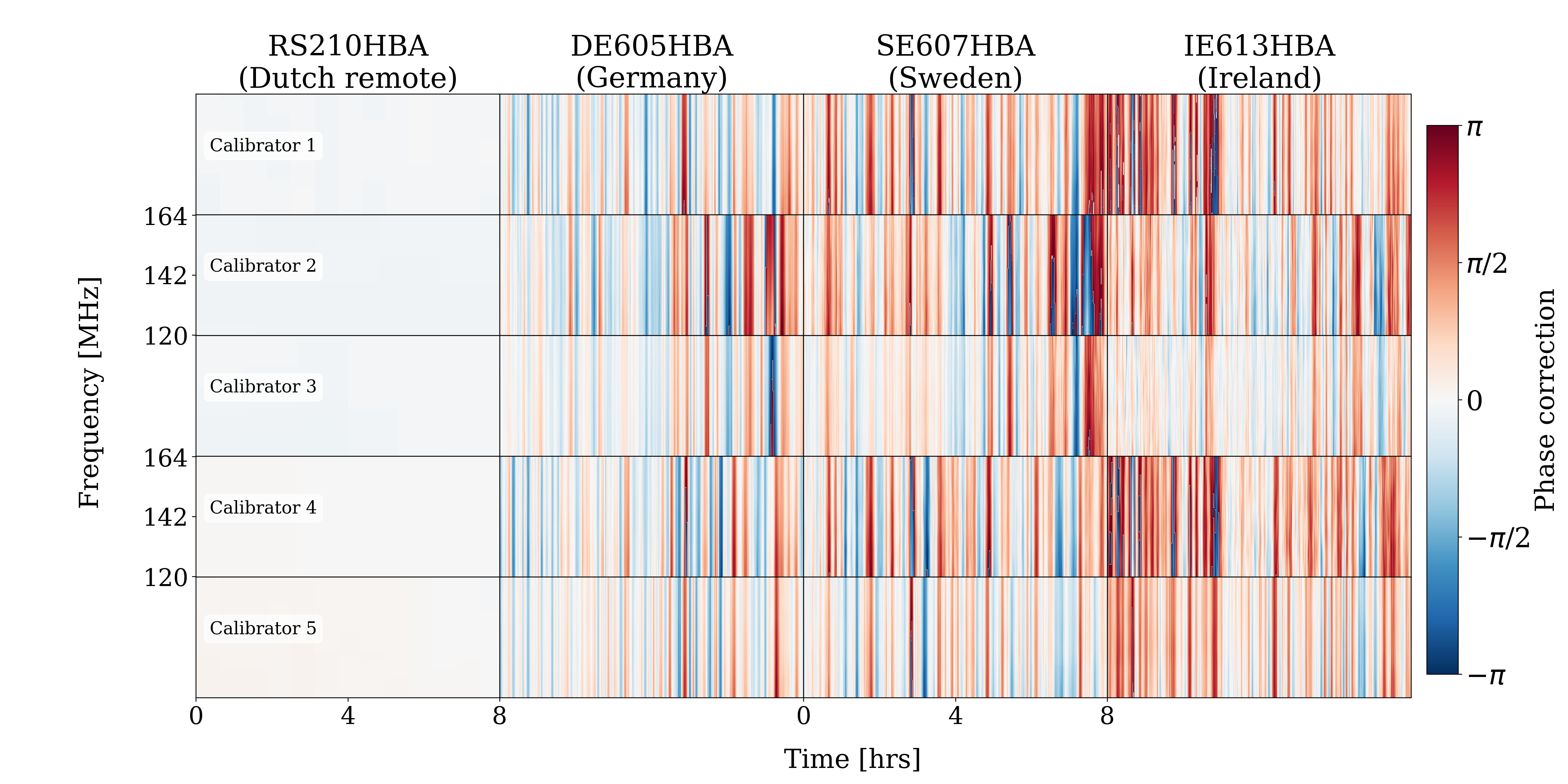}
\includegraphics[width=0.99\linewidth]{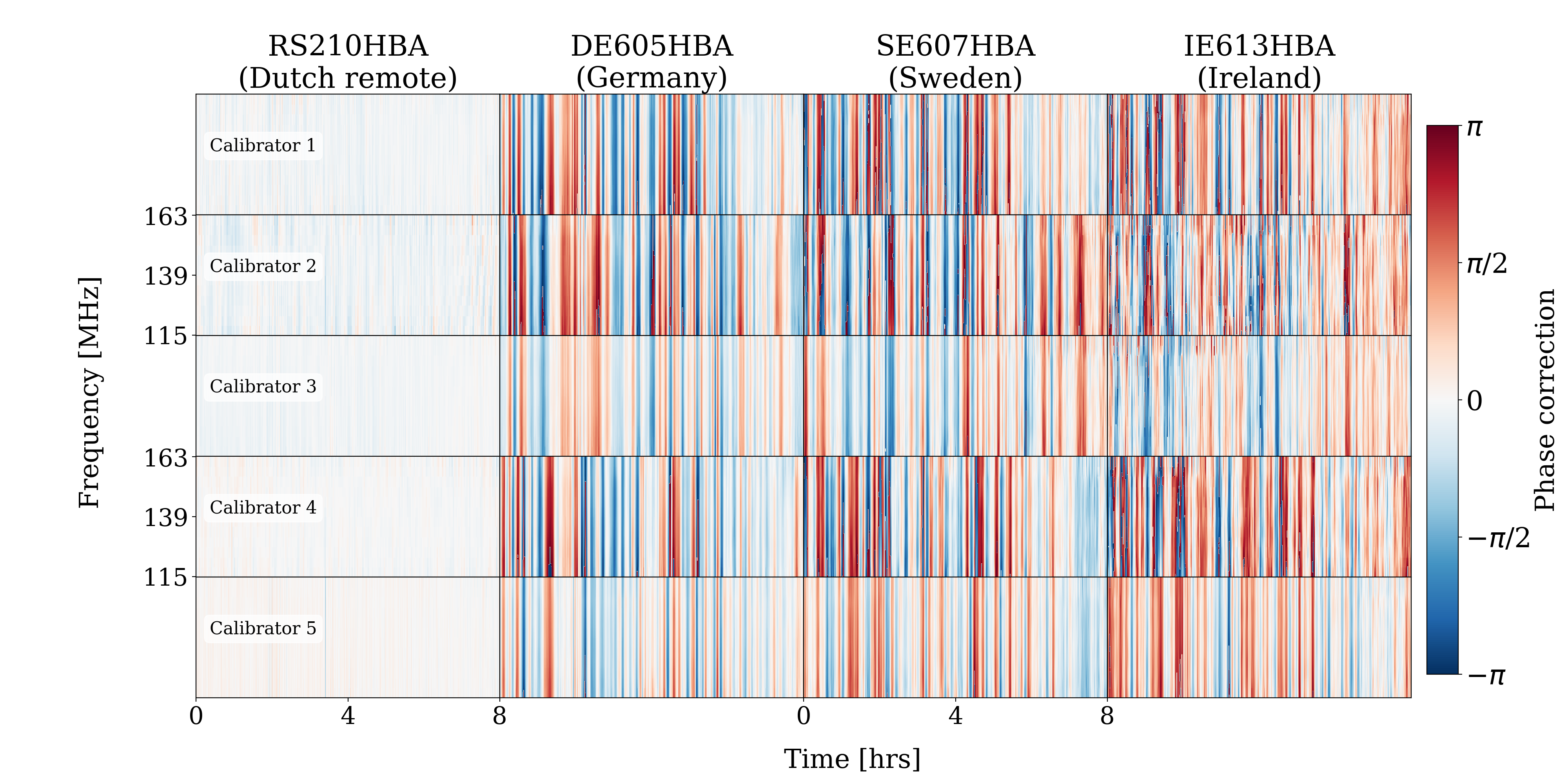}
  \caption{Phase corrections over time and frequency for the five calibrators from Figure \ref{fig:selfcal_comparison} and for four different LOFAR stations, which are positioned as indicated in Figure \ref{fig:stationlayout}. The \textit{upper} panel corresponds to observation \texttt{L686962} and the \textit{lower} panel corresponds to observation \texttt{L833466}.}
\label{fig:selfcal_comparison_solutions}
\end{figure*}

We selected two LOFAR observations of the ELAIS-N1 field that reflect contrasting ionospheric conditions (see Section \ref{sec:data}), where one was taken under relatively calm conditions (\texttt{L686962}) and the other during stronger ionospheric disturbances (\texttt{L833466}). To assess the robustness of our calibration strategy under these different conditions, Figure \ref{fig:selfcal_comparison} shows post-calibration images for five different calibrator sources from both observations, with their corresponding phase corrections displayed in Figure \ref{fig:selfcal_comparison_solutions}.

The calibrator sources generally exhibit higher S/N in \texttt{L833466} than in \texttt{L686962}, mainly due to a \textasciitilde10\% increase in background noise for images from observation \texttt{L833466}. An exception is calibrator source 2, for which the S/N remains comparable between the two observations. This is because the elevated noise in \texttt{L686962} is counterbalanced by time smearing in \texttt{L833466}, which reduces the peak intensity. This source lies closer to the edge of the field, where smearing effects are more pronounced in \texttt{L833466} (see Section \ref{sec:data}).

Residual artefacts are more evident in the images from \texttt{L833466}, as expected from the noisier phase corrections seen in Figure \ref{fig:selfcal_comparison_solutions}. This figure also highlights the difference in phase behaviour for the Dutch remote station RS210, which has stable corrections for data from \texttt{L686962} and considerably wilder corrections for data from observation \texttt{L833466}. Among the calibrators, source 5 appears in image space to be the most problematic in \texttt{L833466}, which is not unexpected, as it has the highest S/N after the in-field calibrator. Despite the quality difference between the observations, the phase corrections in both observations appear stable for this source.

\subsubsection{Image quality refinements}\label{sec:refinements}

\begin{figure}
\includegraphics[width=1\linewidth]{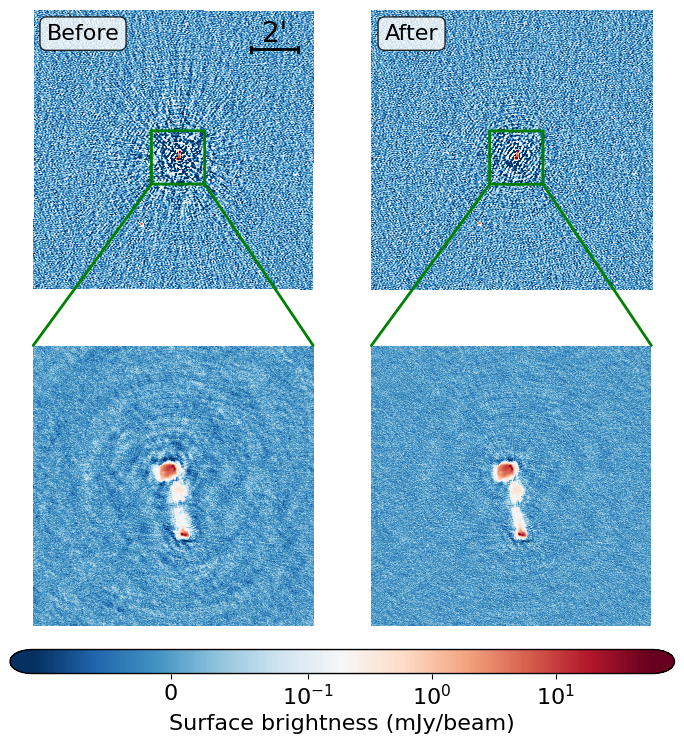}
  \caption{Final image quality improvements at 1.2\arcsec~resolution (\textit{upper} panel) and 0.3\arcsec~resolution (\textit{lower} panel), after joint-calibration with two observations. The \textit{left} panels display the image quality before refinement (up to the calibration from Section \ref{sec:internationalcal}), while the \textit{right} panel presents the image after refinement (including calibration discussed in Section \ref{sec:selfcal2}). We included a 2\arcmin~scale bar to illustrate the size of the image.}
\label{fig:improved}
\end{figure}

With better source subtraction before splitting out datasets for individual facets of our full mosaic, we mitigate the negative effect of high S/N sources affecting the self-calibration of the calibrators corresponding to each facet. This allows us, as outlined in Section \ref{sec:selfcal2}, to perform a final joint self-calibration refinement step, combining both observations.

We tested the final calibration step for the full European array, with and without phasing up the Dutch core stations. For some facets, we found that the phase-up was not required, as most of the bright calibrator sources had been removed from the data, minimising their impact on the calibration of shorter baselines. However, in a few cases, sources within the facet, other than the calibrator source, contained enough signal to still disrupt the calibration of short baselines. Therefore, in particular for automated approaches, it may be recommended to include a phase-up of the Dutch core stations to ensure robust calibration in the final refinement step. This also speeds up the calibration \citep{morabito2022}.

Figure \ref{fig:improved} illustrates the image quality enhancements that we achieve with the final refinement step, for calibrating one of the most challenging high S/N calibrator sources in the ELAIS-N1 field. This source corresponds to calibrator source 5 in Figure \ref{fig:selfcal_comparison}, which experienced the most residual artefacts and also introduced the most artefacts across the full wide-field image from \cite{dejong2024}. At 1.2\arcsec~resolution, most of the radial spike-like artefact structure originating from the calibrator source is noticeably reduced from the left to right panel, though some residual artefacts remain. At 0.3\arcsec~resolution, the improvements are more pronounced, with most artefacts appearing to be fully mitigated. It is expected that the 0.3\arcsec~resolution images exhibit fewer artefacts than the 1.2\arcsec~ones, as LOFAR has \textit{uv}-sampling gaps between 80 and 180~km \citep[e.g.][]{dejong2024} and given that our final joint-calibration strategy is focused on calibrating the longer baselines, using large inner \textit{uv}-cuts. However, the fact that both image resolutions show improvements demonstrates that performing a final calibration step after facet subtraction can further enhance overall calibration quality. In this case, we also find that the residual artefacts from calibrator source 5 in Figure \ref{fig:selfcal_comparison} are largely mitigated during the joint-calibration.

\subsection{Computational cost}\label{sec:cpucost}

The computing costs reported in \cite{dejong2024} summed to 680,000~CPU~core~hours for calibrating and imaging 32~hours of data. This was determined by multiplying the number of CPU cores by the wall-clock time in hours. In this section, we adopt the same method to ensure a consistent comparison. Since half of the data from \cite{dejong2024} was before data reduction already averaged to 2~sec, this corresponds to about 230,000~CPU~core~hours for processing one 8~hour observation at 1~sec time resolution.
With the strategy advancements presented in this work, we have reduced these high costs, as outlined in this subsection.

\begin{figure*}
\includegraphics[width=0.49\linewidth]{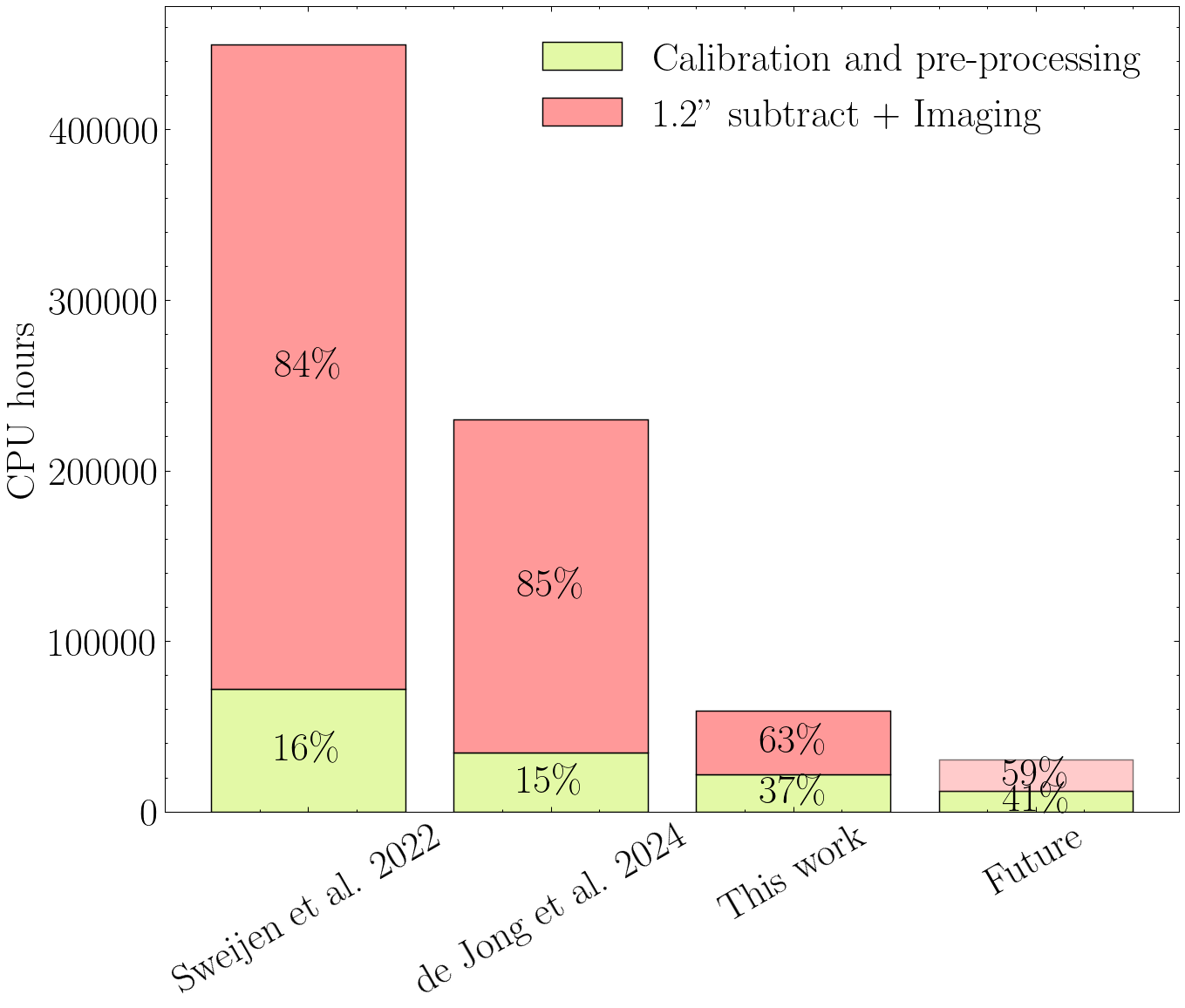}
\includegraphics[width=0.49\linewidth]{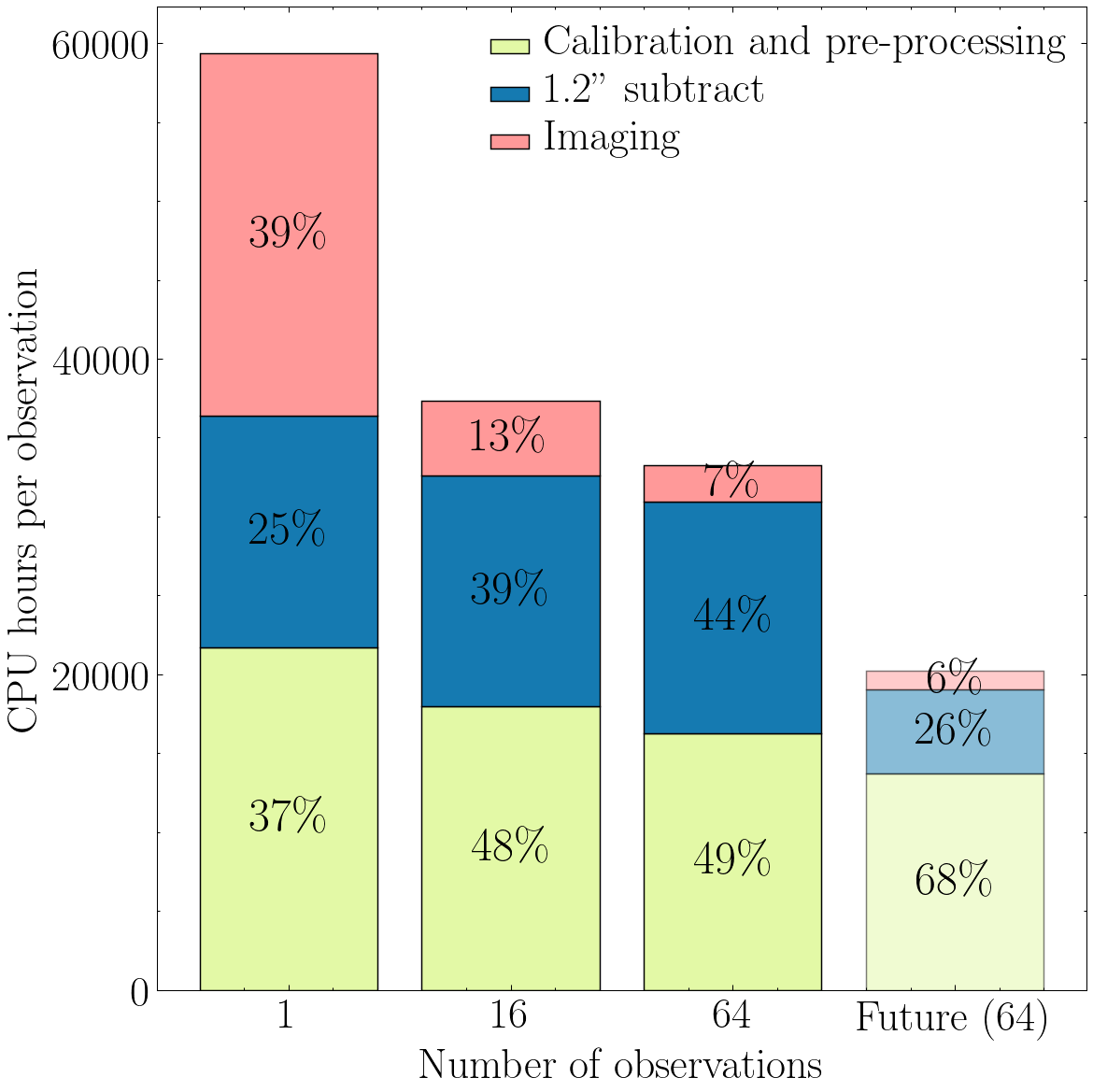}
\caption{Breakdown of the computational cost across the major data processing steps. \textit{Left panel:} comparison between the works by \protect\cite{sweijen2022}, \protect\cite{dejong2024}, and this work for one observation. We also added the potential future computational cost when integrating BDA in both time and frequency during the 1.2\arcsec~imaging and subtraction step and after optimising the calibration strategy (see Section \ref{sec:futurecomp}). \textit{Right panel:} Comparing the computational cost when processing multiple 8~hour observations using SVA with the current strategy and the future strategy with 64 observations. The reduced calibration cost is achieved by using models from previously calibrated observations to speed up the calibration of new observations. We separated the computational cost for calibration from imaging and the 1.2\arcsec~subtraction.}
\label{fig:barchart}
\end{figure*}

A new processing step presented in this work is the additional self-calibration step for the Dutch core and remote stations (see Section \ref{sec:dutchcal}). This increases the total computational costs for every observation by about 8,000 CPU~core~hours. However, for ultra-deep imaging, this cost can be partly mitigated by applying the Dutch calibration for a few observations to generate a sufficiently deep model, which can then be used as input for calibrating the remaining observations. This approach reduces the number of self-calibration cycles required for subsequent observations, lowering the overall cost by a factor of \textasciitilde3 compared to self-calibration without an initial model. Alternatively, revisiting the calibration strategy prior to this step could make it possible to replace this new step entirely by using the DD calibration solutions from the \texttt{DDF}-pipeline (or alternative Dutch LOFAR DD-calibration pipeline), as we discuss in Section \ref{sec:futurecomp}.

Employing early-stopping with the neural network from Section \ref{sec:earlystopping} allows us, in most cases, to already reach convergence between 6 to 8 self-calibration cycles. This reduces the overall computational cost for this calibration step compared to \cite{dejong2024}. However, the computational cost increases again with the introduction of an additional refinement step from Section \ref{sec:selfcal2}, where we perform another round of self-calibration for all facets (see Section \ref{sec:selfcal2}). Nonetheless, due to recent improvements in speed and data volume handling within \texttt{facetselfcal}, the total computational costs for DD calibration with the full European array remains with about 2,000~CPU~core~hours similar to that of \cite{dejong2024}, accounting for only a few per cent of the overall data processing costs.

The originally most computationally expensive step is the 1.2\arcsec~facet subtraction. While \cite{dejong2024} included the facet subtraction with 1.2\arcsec~models as part of imaging, we have in this work separated this step from the imaging (see Section \ref{sec:subtract}), as in the new strategy it is an essential part before performing the final high-quality self-calibration outlined in Section \ref{sec:selfcal2}. We have significantly reduced the cost for this step by modifying the source subtraction strategy (see Section \ref{sec:subtract}), resulting in a computational cost reduction of a factor 10, bringing it down to approximately 12,000~CPU~core~hours per observation. This is significant given that this step originally accounted for 62\% of the total data reduction. As a result, this enhancement alone speeds up the entire data processing workflow by more than a factor of two. This step could be further reduced by another factor of about two if \texttt{DP3} and \texttt{WSClean} are fully flexible in applying and handling BDA datasets with averaging in both time and frequency throughout the entire workflow.

The final imaging step benefits from SVA for deep imaging (see Section \ref{sec:imaging}), especially when processing a large number of observations. For 128~hours of data, the reduction in CPU~core~hours is approximately a factor of \textasciitilde5, while for 500~hours, the speedup increases to roughly \textasciitilde10, as discussed in detail in \cite{dejong2025}. This makes the computational cost for the final imaging almost negligible compared to the other data processing steps.

Beyond our calibration and imaging strategy changes, we are also benefiting from more efficient usage of computational resources by running our workflows with CWL \citep{amstutz2016, crusoe2022}, Toil \citep{vivian2017}, and SLURM \citep{andy2003}. This allows us to more efficiently distribute our jobs and optimise the number of CPU cores for each sub-step. Given the significant optimisation of computational costs achieved through data processing strategy refinements in this work, profiling the workflows and sub-workflows becomes more essential to further enhance computational efficiency in the near future.

In Figure \ref{fig:barchart}, we compare the total computational costs for reducing LOFAR data with the strategies from \cite{sweijen2022} and \cite{dejong2024} for wide-field 0.3\arcsec~imaging for one observation and the computational costs when using SVA for processing multiple observations for deep imaging.
Using our current strategy and software, we find that processing a single observation requires approximately 60,000 CPU core hours. With SVA for deep imaging, we estimate that this cost will decrease to 37,000 CPU core hours per observation when processing a combined calibrated set of 64 observations \cite{dejong2025}. This demonstrates that, depending on the number of combined observations, the total data processing cost has been reduced by a factor of 7.5 to 12 compared to \citet[][taking into account they averaged the data by a factor 2]{sweijen2022}, and by a factor of 4 to 6 compared to \citet[][taking into account half of their data was averaged by a factor 2]{dejong2024}. The precise factor depends on how many observations are combined together with SVA. We also estimate that the entire data processing will be reduced by about another factor of two when our software is capable of handling BDA in both time and frequency and we further optimise our calibration strategy (as outlined in Section \ref{sec:futurecomp}).

\section{Discussion}\label{sec:discussion}

We have improved the DD calibration and reduced the computational costs of processing a LOFAR dataset for sub-arcsecond wide-field imaging. This enables the scalable production of wide-field images with sub-arcsecond resolution, facilitating the creation of ultra-deep wide-field images with sensitivities on the order of $\upmu$Jy~beam$^{-1}$. This work also marks an important improvement for the development of future automated sub-arcsecond wide-field imaging surveys with the entire European LOFAR. In this section, we will discuss potential improvements to generalise and further enhance the results presented in this work.

\subsection{Parameter robustness}

In Section \ref{sec:earlystopping}, we introduced the use of neural networks to employ early-stopping during self-calibration, which we combined with previously employed solution and image quality metrics, to identify a strategy to automate the DD calibration in Section \ref{sec:internationalcal}.
Although we have shown in Section \ref{sec:calqualimpr} that the parameters in this strategy work well for ELAIS-N1 datasets with both wild and calm ionospheres, it is for future automated surveys to consider the robustness of these settings.

Our calibrator source selection resulted in 23 strong and 8 weaker sources (see Section \ref{sec:selection}). Using these across two observations, we were able to optimise calibration parameters under a range of S/N levels and different ionospheric conditions. Compared to other recently imaged LOFAR deep fields \citep[e.g.][Escott et al. in prep.; Bondi et al. in prep.]{sweijen2022}, the calibrator sources in our field show similar properties in both structure and S/N. More than half of the calibrators are point sources or other highly compact objects, while others are (slightly) resolved objects, such as FRII-type radio galaxies with compact lobes \citep{fanaroff1974}, as shown in the examples in Figures \ref{fig:pred_exs} and \ref{fig:selfcal_comparison}. We therefore expect our calibration strategy to be effective in those fields as well, although further testing is required to confirm this. However, given the limited number of datasets we have processed for sub-arcsecond wide-field imaging with LOFAR, there is likely still merit in further refining the calibration parameters, as we continue to reduce additional datasets. This could involve extending the grid-search to a larger and more diverse sample of calibrators drawn from multiple LOFAR observations, especially those with more complex and challenging calibrator sources, lower declinations, or more extreme ionospheric conditions. Doing so will help to generalise the automated DD calibration strategy.

In cases where self-calibration continues to fail, such as possibly in more unique sky areas where calibrators are located within complex dense galaxy clusters \citep[e.g.][]{derubeis2025}, a final validation step may be essential to provide feedback to users when automatically calibrating their data. If the feedback indicates bad calibration performance, users can then intervene and manually adjust the parameters to enhance calibration quality. These validation results can also be valuable for developers, helping to continuously improve the automated calibration and enhance its robustness.

\subsection{Advanced calibration strategies}

In the DD self-calibration images from Figure \ref{fig:pred_exs}, we observe that while the self-calibration results in the last two rows show minor improvements after the initial cycles, additional cycles do not lead to full convergence. This demonstrates that there is still room to improve the calibration.

One way to advance our current strategy is by automatically adjusting the calibration parameter settings during self-calibration. A possible approach is to begin with a conservative set of parameters in the first self-calibration cycle, then gradually shift to less restrictive settings during the following self-calibration cycles, continuing until convergence is achieved according to our metrics. This can include adjustments to shorter solution intervals and reduced frequency smoothness constraints. If self-calibration starts to diverge, measured by assessing the noise on the solutions, parameters can be adjusted to more conservative values.

\begin{figure*}
\centering
\includegraphics[width=0.8\linewidth]{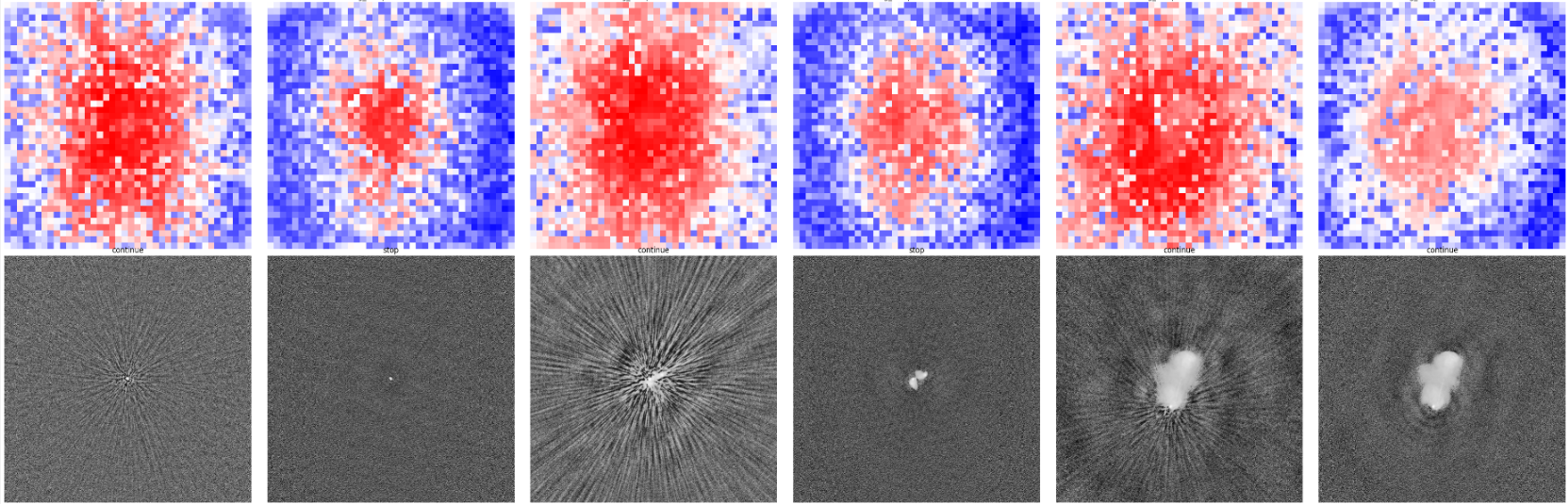}
  \caption{Visualization of the final embedding for the individual patches in the feature extractor (top row) for both before and after calibration for three different sources that we imaged in gray scale (bottom row). The visualization is obtained by performing dimensionality reduction through PCA on the final patch embeddings to obtain a single value that describes the presence of the most prominent feature. Every column corresponds to the same source.}
\label{fig:nn_attention2}
\end{figure*}

To understand how the neural network from Section \ref{sec:earlystopping} interprets the self-calibration images, we visualise the model's patch embeddings in Figure \ref{fig:nn_attention2}, using principal component analysis (PCA). This figure indicates that for images containing sources with strong artefacts, the model primarily focuses on encoding these artefacts, whereas for images that have fewer artefacts, its attention shifts towards the central regions where we find the calibrator source. These images also indicate that there is information about the source and artefact structure being captured by our neural network. An AI model trained to distinguish different calibration artefacts and source structures could be used to tailor calibration parameters, by predicting the optimal settings after each self-calibration cycle. Such an approach could improve both the automation and optimisation of the calibration process.

Looking further ahead, more advanced approaches such as reinforcement learning could be used for automated parameter tuning as well \citep[e.g.][]{yatawatta2021, kirk2025}. This could be achieved without generating images by for instance defining a loss function based on the least-squares difference between corrected and model visibilities after each self-calibration cycle. However, training such models has so far only been demonstrated on smaller datasets due to the high computational costs. This highlights the need for GPU-accelerated implementations of calibration algorithms to reduce the associated costs.

\subsection{Future computational cost reductions}\label{sec:futurecomp}

With our data reduction strategy enhancements, we have reduced the computing costs by a factor of 4 to 6 compared to \citet{dejong2024}, depending on how many LOFAR observations are combined. Nonetheless, more gains in terms of computing costs and I/O handling are still possible through further strategy optimisations, but also software and data volume improvements.

The calibration of the Dutch LOFAR stations happens currently both during the \texttt{DDF}-pipeline and after the DI in-field calibration of the Dutch and international stations with \texttt{facetselfcal} (see Section \ref{sec:dutchcal}). However, it may be more efficient to allow the option to pre-apply the DD solutions from the \texttt{DDF}-pipeline before performing the DD calibration of all European LOFAR stations together. This implies that the \texttt{DDF}-pipeline should run after the DI in-field calibration in the strategy outlined by \cite{sweijen2022} and \cite{dejong2024}. This makes the additional DD calibration step for the Dutch stations with \texttt{facetselfcal} (from Section \ref{sec:dutchcal}) unnecessary. This may lead to computing cost reductions of up to 8,000 CPU~core~hours over the entire data processing. However, before adopting this revised strategy, we should first verify that reordering it does not affect the resulting image quality.

The reordering step in \texttt{WSClean} is currently missing support for frequency-domain BDA. However, this limitation is purely due to the software implementation and is expected to be addressed in the near future. Once supported, it should enable data storage reductions and computational speed-ups comparable to those realised with time-domain BDA (see Section \ref{sec:bda}).
Combining BDA in both the time and frequency domains with lossy compression offers the potential to reduce our original 1~sec, 12~kHz resolution datasets from the original 16~TB to about 1~TB or even lower, depending on the compression level. In addition to lowering long-term data storage costs, this would also significantly reduce I/O overhead. Together with the proposed calibration strategy enhancement outlined above, this could reduce the total computational cost of processing a single 8-hour observation for sub-arcsecond wide-field LOFAR imaging from \textasciitilde60,000 CPU~core~hours to below 30,000 CPU core hours. This is an improvement of about a factor of 8 compared to \cite{dejong2024} and a factor of 15 compared to \cite{sweijen2022}.

Combining the suggestions above for ultra-deep imaging with SVA, the total computational cost will be even further reduced to about 20,000 CPU~core~hours per observation, when for instance 64 observations are combined. This gives a reduction of the total computational costs by a factor of 12 compared to \cite{dejong2024} and a factor of 22 compared to \cite{sweijen2022}. The reduced cost is also indicated in Figure \ref{fig:barchart} with the shallower `future' bar.

By developing and adopting GPU-based alternatives for some of the CPU-intensive tasks in our data processing strategy, we can potentially accelerate several processing steps in the foreseeable future. Porting (parts of) the imaging steps to GPUs has been demonstrated in some cases to give reductions in wall-clock time \citep[e.g.][]{veenboer2020, gheller2023, rubeis2025}. However, a key challenge remains the handling of large data volumes in the case of sub-arcsecond wide-field imaging. As briefly discussed in Section \ref{sec:automated} in the context of reinforcement learning, we also expect that GPU implementations of self-calibration solvers will yield substantial performance gains. This has, for example, been demonstrated for The PetaFLOP AARTFAAC Data-Reduction Engine (PADRE) project, where a speedup of a factor of 6.5 after porting a CPU implementation of a DD calibration algorithm (\texttt{DDECAL}) to GPUs was achieved \citep{vdhaak2024}. While the use of GPUs can reduce wall-clock time, it is important to consider their potential for improved energy efficiency compared to CPU implementations as well \citep[e.g.][]{kruithof2023, lacopo2025}. This makes GPU implementations not only a practical choice for faster generation of science-ready products, but also an important step toward sustainable computing.

\section{Conclusions}\label{sec:conclusion}

In this work, we addressed key challenges that limit the processing of large numbers of observations for sub-arcsecond wide-field imaging with LOFAR. Our main improvements include:
\begin{itemize}[leftmargin=*]
\item An automated DD calibration strategy, with a focus on improving the calibration solutions for the Dutch stations and removing the need for human interventions, by combining already existing image and solution quality metrics with a neural network that assesses image quality as well.
\item A strategy for reducing the computing costs of wide-field facet-based imaging by a factor of 4 to 6 compared to the most recent similar work. The exact factor depends on using one or combining tens of LOFAR observations with SVA.
\end{itemize}
We validated our new data reduction strategy using two observations from the ELAIS-N1 deep field, where one was taken under relatively calm ionospheric conditions and the other during more disturbed conditions, providing us with a robust test of the strategy's performance across varying ionospheric environments. We find clear improvements for in particular the calibration of the shorter baselines, compared to previous work, which enhances the quality for both the 1.2\arcsec~and 0.3\arcsec~resolutions. 

We have also identified several steps to further enhance the data reduction strategy and reduce the total computational cost:
\begin{itemize}[leftmargin=*]
\item We suggest putting efforts into advancing the methods to perform automated parameter-tuning. This may involve adjusting parameters between different self-calibration cycles, using for example, image-based clustering methods to incorporate information about source or artefact structure from uncalibrated images, or perhaps by adopting reinforcement learning. This could be particularly beneficial for the most challenging, complex high S/N sources, where our current method still leaves room for improvement.
\item To reduce the cost for calibration, we could reorder some processing steps and perform DI in-field calibration before running the \texttt{DDF}-pipeline or \texttt{facetselfcal} for joint-DD calibration of the Dutch stations. This implies that we only perform DD calibration for the Dutch stations once, which may eventually save an additional 13\% on total computing time.
\item Enabling full support in \texttt{DP3} and \texttt{WSClean} for calibration and imaging datasets with BDA in both time and frequency domains applied, could reduce the computational cost by more than a factor of two compared to this work. Furthermore, combining BDA with Dysco compression would also significantly reduce data storage requirements, which will be essential for wide surveys. Porting parts of the CPU-intensive tasks to GPU-based implementations may be another way to lower both wall-clock time and energy costs.
\end{itemize}
In our forthcoming work (de Jong et al. in prep.), we will apply the calibration and imaging strategy from this work to 200 hours of ELAIS-N1 data, aiming to reach sensitivities on the order of 6~$\upmu$Jy~beam$^{-1}$. This will demonstrate that the presented techniques in this work allow us to make ultra-deep high-resolution images with LOFAR. Our work also marks a key step toward establishing a fully automated survey workflow for sub-arcsecond imaging with LOFAR.

\section*{Acknowledgements}

This work is part of the project CORTEX (NWA.1160.18.316) of the research programme NWA-ORC which is (partly) financed by the Dutch Research Council (NWO). This work made use of the Dutch national e-infrastructure with the support of the SURF Cooperative using grant no. EINF-1287. This work is co-funded by the EGI-ACE project (Horizon 2020) under Grant number 101017567.
The authors acknowledge the OSCARS project, which has received funding from the European Commission’s Horizon Europe Research and Innovation programme under grant agreement No. 101129751.
RJvW acknowledges support from the ERC Starting Grant ClusterWeb 804208. 
LKM is grateful for support from a UKRI FLF [MR/Y020405/1] and LOFAR-UK through STFC [ST/V002406]. 
The work of MvdW was supported by the Science and Technology Facilities Council via LOFAR-U.K. [ST/V002406/1]. 
The use of the national computer facilities in this research was subsidised by NWO Domain Science.
LOFAR data products were provided by the LOFAR Surveys Key Science project (LSKSP; \url{https://lofar-surveys.org/}) and were derived from observations with the International LOFAR Telescope (ILT). LOFAR \citep{haarlem2013} is the Low-Frequency Array designed and constructed by ASTRON. It has observing, data processing, and data storage facilities in several countries, which are owned by various parties (each with their own funding sources), and which are collectively operated by the ILT foundation under a joint scientific policy. The efforts of the LSKSP have benefited from funding from the European Research Council, NOVA, NWO, CNRS-INSU, the SURF Co-operative, the UK Science and Technology Funding Council and the Jülich Supercomputing Centre.

\section*{Data availability}

This paper presents data reduction and analysis techniques without publishing new observational data. The methods are demonstrated using publicly available data from the LOFAR Long-Term Archive (LTA), accessible at \url{https://lta.lofar.eu}, which was already partly processed by \cite{dejong2024}. The workflow described in this work will be incorporated into the LOFAR VLBI pipeline, available at \url{https://git.astron.nl/RD/VLBI-cwl}.

\bibliographystyle{mnras}
\bibliography{bib}

\appendix

\section{Phasediff-score and S/N}\label{sec:phasediff_snr}

\begin{figure}
\centering
\includegraphics[width=0.95\linewidth]{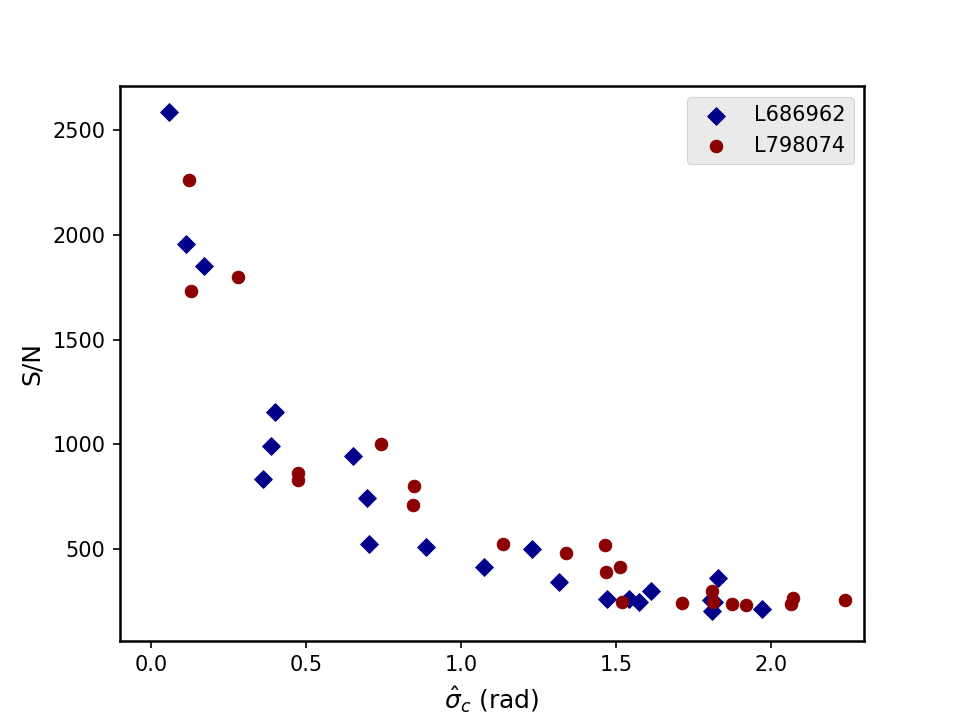}
  \caption{Comparison between the phasediff-score and S/N of different calibrator sources and across the two observations analysed in this work. The S/N is computed after calibrating and imaging the calibrator sources, following the procedure in Section \ref{sec:internationalcal}, and is defined as the ratio of peak intensity to background RMS noise in the resulting images.}
\label{fig:phasediff_snr}
\end{figure}

To demonstrate that the phasediff-score ($\hat{\sigma_{c}}$) traces S/N, we display in Figure \ref{fig:phasediff_snr} the S/N as a function of the phasediff-score for various calibrator sources across both observations analysed in this study. The strong inverse relationship between the two is reflected in a Spearman rank correlation coefficient of $−0.94$ for observation \texttt{L686962} and $-0.92$ for observation \texttt{L833466}, confirming that the phasediff-score serves as a reliable proxy for S/N under both calm and disturbed ionospheric conditions. Note also in this figure that the lowest phasediff-scores correspond to the in-field calibrator source, which was for ELAIS-N1 originally selected by \cite{ye2024}.

\section{Circular statistics}\label{sec:circstat}

The circular standard deviation is given by
\begin{equation*}
\sigma_{c} = \sqrt{-2 \ln{R}},
\end{equation*}
where $R$ is the resultant length, which is a value between 0 and 1 and captures information about the distribution of a set of angles (i.e. phase corrections in our case). Values near 0 indicate widely dispersed angles, whereas values near 1 indicate angles that are very similar. The mean resultant is generally defined as \citep[e.g.][]{mardia1972}:
\begin{equation*}
R=\sqrt{\bar{C}^{2}+\bar{S}^{2}},
\end{equation*}
with the mean cosine angles
\begin{equation*}
\bar{C}=\frac{1}{N}\sum_{i=1}^{N}\cos{\theta_{i}}
\end{equation*}
and the mean sine angles
\begin{equation*}
\bar{S}=\frac{1}{N}\sum_{i=1}^{N}\sin{\theta_{i}}.
\end{equation*}
We find with Equation \ref{eq:circscore} that the resultant is in function of solution interval $\delta_{t}$ equal to
\begin{equation}\label{eq:rt}
R(t) = \exp{\left[-\frac{\pi^{2}}{2}\left(1-\exp{\left(-\frac{\varsigma}{\sqrt{\delta_{t}}}\right)}\right)\right]}.
\end{equation}
This means that for $\delta_{t} \rightarrow 0$ we find
\begin{equation*}
R \rightarrow \exp{\left(-\frac{\pi^{2}}{2}\right)}\approx 0.0072,
\end{equation*}
while for $\delta_{t} \rightarrow \infty$ we find
\begin{equation*}
R \rightarrow \exp{(0)}=1.
\end{equation*}
This aligns with the fact that shortening the solution intervals increases the variance of the phase corrections, since this reduces the S/N within a solution interval, whereas enlarging solution intervals averages out the variance of the phase corrections. The challenge is to strike a balance between selecting sufficiently short solution intervals and maintaining an acceptable level of phase correction variance in order to accurately capture corrections over the observation without self-calibration divergence. This is outlined in Sections \ref{sec:solints} and \ref{sec:internationalcal}.

We can also derive the circular variance, which is according to the theory of circular statistics equal to
\begin{equation*}
V_{c} = 1 - R.
\end{equation*}
Inserting equation \ref{eq:rt} gives us
\begin{equation*}
V_{c} = 1 - \exp{\left[-\frac{\pi^{2}}{2}\left(1-\exp{\left(-\frac{\varsigma}{\sqrt{\delta_{t}}}\right)}\right)\right]},
\end{equation*}
from which we can estimate with a Taylor expansion that as $\delta_{t}\gg\varsigma^2$, we get
\begin{equation*}
\sigma_{c}^2 \rightarrow 2V_{c}.
\end{equation*}
This agrees with the theory of circular statistics for small $\sigma_{c}$.

\section{Calibration software}\label{sec:softwaredepen}

The calibration steps outlined in Section \ref{sec:ddcal} utilise \texttt{facetselfcal}\footnote{\url{https://github.com/rvweeren/lofar_facet_selfcal}} \citep{vanweeren2021}. This is a flexible low-frequency self-calibration software package, which integrates the Default Preprocessing Pipeline (\texttt{DP3}; \citealt{dp3, dijkema2023}) and \texttt{WSClean} \citep{wsclean} for (self-)calibration. \texttt{facetselfcal} executes multiple self-calibration cycles, adjusting the source model iteratively before proceeding to the next cycle. 

Since our calibration operates automatically, we implemented in \texttt{facetselfcal} new tools that offer quick insights into its parameter tuning. This aligns with broader initiatives to enhance the findability, accessibility, interoperability, and reproducibility \citep[FAIR; e.g.][]{wilkinson2016, otoole2022} of data and pipelines for advancing open science in LOFAR long-baseline processing. These functionalities embed metadata containing information about the adjusted parameters into the output calibration solutions, enabling users to manually re-run the calibration with modified settings if desired (advancing reproducibility).

Next to \texttt{facetselfcal}, we also make use of \texttt{lofar\_helpers}\footnote{\url{https://github.com/jurjen93/lofar_helpers}}. This software package was introduced by \cite{dejong2022} and provides tools on top of \texttt{DP3}, \texttt{casacore}\footnote{\url{https://casacore.github.io/python-casacore/}} \citep{casacoreteam, casateam2022}, and \texttt{astropy}\footnote{\url{https://www.astropy.org}} \citep{astropy1, astropy2, astropy3} to perform various dataset and image manipulations.

\section{Modified DINOv2 neural network model}\label{sec:nn_descript}

We introduce in Section \ref{sec:earlystopping} the use of a pre-trained neural network to classify calibration images, which we have fine-tuned using self-calibration images from our ELAIS-N1 data and other datasets. In this Appendix section, we will provide more details about the pre-trained model we have used and how we have adjusted it to our own needs. 

We use an already pre-trained 2nd version of the self-distillation with no labels model \citep[DINOv2;][]{oquab2023}, utilising register-based methods \citep{darcet2023} with a vision transformer \citep[ViT;][]{dosovitskiy2020} as the backbone. The model we use has a ViT-B backbone, comprising 86 million parameters, which was trained using knowledge distillation \citep{hinton2015distilling} on a larger DINOv2 model with a ViT-g backbone containing 1.1 billion parameters. The DINOv2 pre-training dataset (LVD-142) contained a combination of existing curated datasets, including different ImageNet datasets \citep{russakovsky2015}, Google Landmarks \citep{weyand2020}, and images scraped from the internet. The total combined dataset consisted of 142~million images. The original DINOv2 model was trained in a self-supervised manner on this comprehensive dataset.

We have taken the already pre-trained DINOv2 model and only substituted the classification sub-model with a custom classifier. This classifier is a two-layered perceptron \citep{rumelhart1986}. 
In this classifier, a dropout is applied during training to the feature extractor's output for regularisation. This prevents the model from focussing too much on a few features, and thus prevents overfitting. Subsequently, two linear layers are applied: the first layer, activated by a rectified linear unit \citep[ReLU;][]{nair2010}, maps the feature extractor's output to the same dimensionality of the output. The second layer reduces this output to a single value, followed by a sigmoid function to produce a pseudo-probabilistic output.
For fine-tuning the DINOv2 feature extractor we employ Low-Rank Adaption \citep[LoRA;][]{hu2022lora}, where the updates for each weight matrix are parametrised by two low-rank matrices with a tunable rank $r$.\footnote{Given a weight update $\Delta W\in \mathbb{R}^{m\times n}$ and rank $r$, the low-rank decomposition is given by the product of $\Delta W_1 \in \mathbb{R}^{m\times r} $ and $\Delta W_2\in \mathbb{R}^{r\times n}$.} Reducing the rank reduces the effective number of trainable parameters, improves compute performance and memory efficiency, and reduces the flexibility of the model during fine-tuning, reducing the risk of over-fitting and improving generalisation. In our case, we use a rank of $r=32$. This choice balances performance and computational cost.
To further improve regularisation we apply label smoothing as well \citep{szegedy2015}. This is a technique that slightly reduces the \textit{continue} class $P=1$ to, for instance, $P=0.9$ and adds the remaining probability (e.g., 0.1) to the \textit{stop} class. This reduces overconfidence in the model's predictions, particularly for noisy target labels, helping it generalise better by not focusing too strictly on exact class labels. Our implementation built on the DINOv2 model is available in our open-source repository.\footnote{\url{https://github.com/LOFAR-VLBI/astroNNomy}}

The effectiveness of utilising DINOv2 as a pre-trained base model has been demonstrated across various domains, including applications in medical imaging \citep[e.g.][]{Kundu2024,song2024} and geological imaging \citep[e.g.][]{Brondolo2024}. This is because the original DINOv2 model was primarily trained on generic natural images sourced from the internet, and its self-supervised training scheme enables the extraction of more general and broadly applicable features than those learned through supervised training \citep[e.g.][]{huang2021}.
The primary advantage of transfer learning from the DINOv2 model is also that, as long as the data modalities remain relatively similar, the extracted features are likely to be transferable \citep[e.g.][]{gerace2022,tahir2024}. 
This allows us to achieve high performance by utilising the well-trained feature extractor of the DINOv2 model, which is pre-trained on a large dataset, rather than training the feature extractor from scratch on a smaller dataset. By training a classifier that acts on these features and fine-tuning the feature extractor on our domain-specific images, we can obtain a model that is robust for detecting artefacts in these images. This can be assessed by evaluating the model on test images that are unseen during the training stage.

\label{lastpage}
\end{document}